\documentclass[apj]{emulateapj}
\usepackage{epsfig}
\usepackage{amsmath}
\usepackage{amssymb}
\usepackage{natbib}
\usepackage{setspace}
\usepackage{graphicx}

\newcommand{\degree}{\ensuremath{^\circ}}
\newcommand{\kmskpc}{\ensuremath{km s^{-1} kpc^{-1}}}

\begin{document}
\title{Gas Kinematics on GMC scales in M51 with PAWS: cloud stabilization through dynamical pressure }
\author{Sharon E. Meidt\altaffilmark{1}, Eva Schinnerer\altaffilmark{1}, Santiago Garc\'{i}a-Burillo\altaffilmark{2}, Annie Hughes\altaffilmark{1}, Dario Colombo\altaffilmark{1}, J\'er\^ome Pety\altaffilmark{3}$^,$\altaffilmark{4}, Clare L. Dobbs\altaffilmark{5}, Karl F. Schuster\altaffilmark{3}, Carsten Kramer\altaffilmark{6}, Adam K. Leroy\altaffilmark{7}, Ga\"elle Dumas\altaffilmark{3}, Todd A. Thompson\altaffilmark{8}$^,$\altaffilmark{9}}

\altaffiltext{1}{Max-Planck-Institut f\"ur Astronomie / K\"{o}nigstuhl 17 D-69117 Heidelberg, Germany}
\altaffiltext{2}{Observatorio Astron\'{o}mico Nacional - OAN, Observatorio de Madrid Alfonso XII, 3, 28014 - Madrid, Spain}
\altaffiltext{3}{Institut de Radioastronomie Millim\'etrique, 300 Rue de la Piscine, F-38406 Saint Martin d'H\`eres, France}
\altaffiltext{4}{Observatoire de Paris, 61 Avenue de l'Observatoire, F-75014 Paris, France.}
\altaffiltext{5}{School of Physics and Astronomy, University of Exeter, Stocker Road, Exeter EX4 4QL, UK}
\altaffiltext{6}{Instituto Radioastronom\'{i}a Milim\'{e}trica, Av. Divina Pastora 7, Nucleo Central, 18012 Granada, Spain}
\altaffiltext{7}{National Radio Astronomy Observatory, 520 Edgemont Road, Charlottesville, VA 22903, USA}
\altaffiltext{8}{Department of Astronomy, The Ohio State University, 140 W. 18th Ave., Columbus, OH 43210, USA} 
\altaffiltext{9}{Center for Cosmology and AstroParticle Physics, The Ohio State University, 191 W. Woodruff Ave., Columbus, OH 43210, USA}
\date{\today}

\begin{abstract}

We use the high spatial and spectral resolution of the PAWS CO(1-0) survey of the inner 9 kpc of the iconic spiral galaxy M51 to examine the effect of gas streaming motions on the star-forming properties of individual GMCs.  
We compare our view of gas flows in M51 -- which arise due to departures from axi-symmetry in the gravitational potential (i.e. the nuclear bar and spiral arms) -- with the global pattern of star formation as traced by H$\alpha$ and 24$\mu m$ emission.  
We find that the dynamical environment of GMCs strongly affects their ability to form stars, in the sense that GMCs situated in regions with large streaming motions can be stabilized, while similarly massive GMCs in regions without streaming go on to efficiently form stars.  
We argue that this is the result of reduced surface pressure felt by clouds embedded in an ambient medium undergoing large streaming motions, which prevents collapse.  
Indeed, the variation in gas depletion time expected based on the observed streaming motions throughout the disk of M51 quantitatively agrees with the variation in observed gas depletion time scale.  
The example of M51 shows that streaming motions, triggered by gravitational instabilities in the form of bars and spiral arms, can alter the star formation law; this can explain the variation in gas depletion time among galaxies with different masses and morphologies. 
In particular, we can explain the long gas depletion times in spiral galaxies compared to dwarf galaxies and starbursts.  
We suggest that adding a dynamical pressure term to the canonical free-fall time produces a single star formation law that can be applied to all star-forming regions and galaxies, across cosmic time.
\end{abstract}

\section{Introduction}

Our understanding of how galaxies build up their mass -- fundamental to our picture of galaxy evolution -- relies heavily on the relation between newborn stars and the dense molecular material out of which they form.  
At present, the empirical Kennicutt-Schmidt (KS; \citealt{schmidt}; \citealt{kennKS}) relation linking the star formation rate surface density $\Sigma_{SFR}$ with a power-law dependence on gas surface density $\Sigma_{gas}$ is in wide use by numerical simulations that inform our view of galaxy evolution.  
This relation has been demonstrated to hold over several orders of magnitude in $\Sigma_{gas}$ and across a range of galaxy types.  
The first homogeneous spatially-resolved observations in nearby galaxies  (e.g. \citealt{bigiel}; Leroy et al. 2008,2009,2012) suggest that gas is consumed in star formation at a fixed efficiency, leading to an apparently `universal', but long, gas depletion time of $\sim$2.5 Gyr for galaxies. 
The relation was further shown to be determined  predominantly by the molecular phase of the neutral gas (\citealt{schruba11}; \citealt{bigiel}).   

At the same time, there is mounting evidence that star formation may occur in two modes: the `normal' mode typical of the majority of star formation in local disks, contrasted by a `starburst' mode, wherein gas is consumed much more efficiently and quickly (i.e. \citealt{kennicutt98}).  
Bimodality in the star formation (SF) relation is manifest both locally (i.e. \citealt{leroy2013}; \citealt{saintonge}) and at higher z (i.e. \citealt{daddi}; \citealt{genzel}; \citealt{gb12}).  But, to date, there is no single `universal' star formation relation than can smoothly link the two regimes, across all times (but see \citealt{krumholz12}).   

How the star-forming gas is organized may influence the depletion time (\citealt{combes}; \citealt{saintonge} and references therein).  
An interpretation of the long depletion time in `normal' star forming galaxies could be that it reflects only the (low) fraction of star-forming clouds rather than the fraction of a cloud that forms stars.  
Variations in the properties of Giant Molecular Clouds (GMCs), the actual seeds of star formation, themselves could profoundly impact their star formation capabilities as evidenced by the large scatter of observed star formation rates (SFRs) in local Galactic GMCs (\citealt{MooneySolomon}; \citealt{kennicuttEvans}; \citealt{heiderman}).   
Yet, our understanding of the molecular gas phase and its organization into self-gravitating entities, i.e. GMCs, is limited (for a review, see \citealt{mckeeostrikerReview}).  

Observations of molecular gas at high resolution in nearby spiral galaxies are critical to test the emerging picture of how galactic
environment influences the organization, properties and evolutionary pathways of a
galaxy's GMC population, and its global patterns of star formation.  In this paper we continue our study of the interacting Whirlpool galaxy M51 (D=7.6 Mpc; \citealt{ciardullo}), which is more representative of a typical star forming galaxy than the well-studied lower mass galaxies in the Local Group.  
The unparalleled high resolution of the PAWS (PdBI Arcsecond Whirlpool Survey; \citealt{schinner2013}) data, combined with exceptional multi-wavelength coverage, makes it an ideal target for 
examining the influence of bar and spiral instabilities -- and more generally, dynamical galactic environment -- on the organization of the ISM and global star formation patterns.  

Our detailed analysis of the molecular gas in M51-- including its GMC population -- so far suggests that several mechanisms control its structure and evolution \citep{colombo2013}.  
Deviations in the gravitational potential from axisymmetry (i.e. the nuclear bar and two-armed spiral) induce shear, shocks, and streaming motions in the gas flow, leading to changes in the local gas surface density.  
This subsequently alters the pattern of radiation from recent star formation, which occurs on the convex side of the spiral arm, and therefore also the mechanical energy input from newly formed OB stars, stellar winds and supernovae.  
The galactic environment is thus very dynamic in both senses, and it is not clear under which conditions the developed static models apply (e.g. Krumholz et al. 2009).  

Gas kinematics on the scales of individual GMCs are key for linking on-going star formation to the organization and radial transport of gas by the stellar spiral arms.  
We expect the dynamical environment to strongly impact the organization of the ISM in a galaxy like M51, which hosts strong, well-organized spiral structure and is presently undergoing interaction with a companion. 
Depending on location in the disk, gas will be driven radially in- or outward and may occasionally be halted along its path, building up rings (see review by \cite{bc}).  
Whether gas is in motion or stationary will naturally influence the build-up of the star-forming gas reservoir, the accumulation of mass in to clouds, and the eventual collapse to form stars. 

To quantify the influence of gas kinematics on cloud scales we undertake a study in two parts.  We begin by evaluating the close link between the gas depletion time $\tau_{dep}$=$\Sigma_{gas}$/$\Sigma_{SFR}$\footnote{In the extragalactic context, the gas depletion time is equated with the inverse of the star formation efficiency SFE=$\Sigma_{SFR}/\Sigma_{gas}$, although the former quantity is not a traditional efficiency, but an efficiency per free fall time.  This is in contrast with the star formation efficiency $\epsilon$ referred to in galactic studies, which represents the fraction of gas converted in to stars per star formation event. } and dynamical environment as defined by the radial profile of present-day gravitational torques ($\S$\ref{sec:torqprof}).  We then establish the magnitude and size of gas motions driven in response to the spiral torquing ($\S$\ref{sec:streaming}) and interpret the behavior in $\tau_{dep}$ in terms of gas kinematics in $\S$ \ref{sec:dynamicalPqual}.   We suggest that dynamical pressure plays a role in stabilizing GMCs.  

Second, we develop a picture that links gas streaming motions to the gas depletion time using the Bernoulli principle, which equates an increase in gas velocity with reduced external cloud pressure ($\S$ \ref{sec:dynamicalPquant}).  
We use the radial profiles of streaming motions to assess the role of shear on the stability of gas and discuss the evidence against shock- and stellar feedback-driven turbulent support in $\S$ \ref{sec:stability}.    
We conclude with a discussion of the broader implications of this result, from the angular offset between gas and young star tracers to the scatter in the KS star formation relation ($\S$ \ref{sec:disc_patterns}).  
In $\S$\ref{sec:disc_SFrelation} we present a new form for the star formation relation between $\Sigma_{SFR}$ and $\Sigma_{gas}$ that smoothly links star formation at low-z and high-z by parameterizing the environmental dependence of clouds in terms of non-circular gas streaming motions.  
In particular, we predict a link between gas depletion time and gas fraction, as recently observed locally \citep{saintonge} and in evolution from z=2 \citep{combes}.

\section{Data}

\label{sec:data}
The analysis in this paper is based on $\mbox{CO (1-0)}$ data obtained by the Plateau de Bure Arcsecond Whirlpool Survey (PAWS,
\citealt{schinner2013}).  
PAWS observations cover a roughly 270\arcsec$\times$170\arcsec ~field of view in the inner disk of M51 and include both interferometric data from the Plateau de Bure
Interferometer (PdBI) and single-dish data from the IRAM 30m telescope.   The effective angular resolution in the final combined data cube is 1.\arcsec16$\times$0.\arcsec97, or a spatial resolution of $\sim$40 pc at our assumed distance of M51.   
The channel separation is 5 km s$^{-1}$.  The PAWS observing strategy, data reduction and combination procedures, and flux calibration are described by \cite{petyPAWS}.   

Here we use maps of the zeroth, first and second moments of the PAWS data cube as presented by \cite{petyPAWS}.  
From the PAWS map of integrated CO(1-0) intensities (moment zero), we estimate the molecular gas surface density, applying a constant conversion factor $X_{CO}$=2$\times$10$^{20}$ cm$^{-2}$ (K km s$^{-1}$)$^{-1}$ (as argued in \citealt{colombo2013}) and including a factor of 1.36 to account for the presence of Helium.  
From the line-of-sight velocity field (first moment) we derive estimates for the circular velocity and the non-circular motions.  
The velocity dispersion map (moment two) supplies our measure of the turbulent gas motions after removal of the  contribution from unresolved bulk motions estimated from the line-of-sight velocity field (adopting the formalism developed by \citealt{petyPAWS}).  We refer the reader to \citet{petyPAWS} for a study of the resolution-dependence of observables such as the maximum CO brightness (i.e. gas surface density) and velocity second moment (gas velocity dispersion) measured from the PAWS data cube. 

\begin{figure}[t]
\hspace*{-.15in}\includegraphics[width=1.1\linewidth]{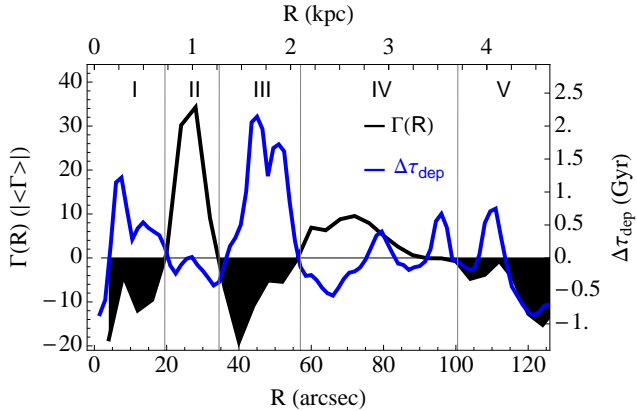}
\caption{Radial profiles of the azimuthally-averaged torque (in units of the absolute magnitude of the average torque across the PAWS field of view $\vert$$<$$\Gamma$$>$$\vert$; black and white) and molecular gas depletion time $\Delta$$\tau_{dep}$=$\tau_{dep}$-$<$$\tau_{dep}$$>$ (where $<$$\tau_{dep}$$>$ is the average $\tau_{dep}$ across the PAWS field of view; blue) measured in 2.4\arcsec ~radial bins.  
Each crossing from negative to positive torque corresponds to the location of the corotation radius (CR) of the structure: inside CR material is driven radially inward and outside material moves outward.  }
\label{fig:torqsfe}
\end{figure}
The PAWS data are complemented by an extensive multi-wavelength data set, which has been assembled and described in \citet{schinner2013}.  
Here we make use of the S$^4$G IRAC 3.6 $\mu m$ image \citep{sheths4g}, corrected for the contribution of non-stellar emission \citep{meidt12}, to define the stellar mass distribution.  
As described in Appendix \ref{sec:appxTorques}, we use this stellar mass map to derive the stellar potential and the profile of present day torques.  
We also use the H$\alpha$ image (corrected for the stellar continuum using the procedure of \citealt{gutierrez}) and MIPS 24 $\mu m$ data acquired as part of SINGS \citep{kennicuttSINGS} to trace the star formation in M51.  The MIPS 24 $\mu m$ image has been processed with the HiRes algorithm, giving a limiting angular resolution of 2.4\arcsec \citep{dumas}. 
We combine the two images to trace the total (obscured and unobscured) star formation following \citet{leroy}, using the empirical calibration of \citet{calzetti07}.  

\section{Gas Flow and Global Star Formation Patterns in M51}
\label{sec:globalSFtorques}

M51 is a favorite test-bed for spiral arm density wave
theories. Studies of the spiral morphology and kinematics show
evidence for the offset alignment of the gaseous, young and old
stellar tracers predicted by theory, as well as strong non-circular
gas streaming motions (\citealt{vog93}; \citealt{gb};
\citealt{rand93}; \citealt{aalto}; \citealt{schuster};
\citealt{shetty}; \citealt{schinner2013}).   In this section, we examine gas motions in M51 using two independent approaches.  We
  calculate the present-day torques on the molecular gas due to the
  stellar component.  We then decompose the CO velocity field, and show that the gas motions implied by our kinematic analysis are in excellent agreement with the response that we would expect based on
  our torque map.   
  Finally, we compare our picture of gas flows with the global pattern of star formation.  

\subsection{Present-day Torques}\label{sec:torqprof}
As described in detail in Appendix \ref{sec:appxTorques}, our 2D map of the old stellar light in M51 \citep{schinner2013} gives us a snap-shot of the torques exerted by the non-axisymmetric structure (i.e. nuclear bar, spiral arms) present in the density and hence gravitational potential of the system.  
These torques drive radially in- and outward gas motions, depending on the sign of torque.  
Crossings from negative to positive in the azimuthally averaged torque profile (Figure \ref{fig:torqsfe}) mark the location of corotation, which is defined as the radius where the angular speeds of the disk and non-axisymmetric structure are the same.  
Inside (outside) corotation, negative (positive) torques drive gas radially inward (outward).  Crossings from positive to negative torque coincide with the switch in predominance to a unique pattern.  

This series of zero-crossings defines a set of dynamical environments (see Appendix \ref{sec:appxTorques}): the zone dominated by the nuclear bar ($R$$\lesssim$20\arcsec), the molecular ring ($R$$\sim$30\arcsec), the inner spiral (30\arcsec$<$$R$$<$85\arcsec) and the outer spiral ($R$$>$85\arcsec).  
The inner spiral can be further split in to two zones, inside and outside the main spiral's corotation radius (CR) at $R$=60\arcsec.  
In what follows, we use this series of dynamical environments -- and the torques within them -- as the basis for interpreting the observed pattern of non-circular gas streaming motions in the inner disk of M51.  

\subsection{Radial and azimuthal streaming motions\label{sec:streaming}}

\begin{figure}[t]
\includegraphics[width=1.09\linewidth]{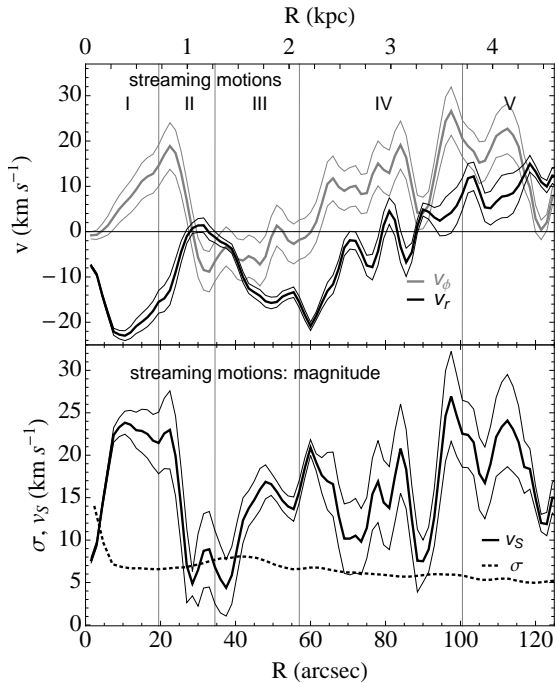}
\caption{(Top) CO intensity-weighted radial (black) and azimuthal (gray) non-circular streaming velocities measured in 1.5\arcsec ~radial bins, reconstructed as described in the text.  
The intensity-weighting focuses our measurements on the CO-bright spiral arms.  Thin lines represent the rms dispersion in solutions comprised of $\pm$5$\degree$ variation in the assumed spiral arm pitch angle $i_p$ added in quadrature with the uncertainty introduced by our adopted rotation curve (see text).  (Bottom) Magnitude of the non-circular streaming velocities $\vert v_S\vert=\sqrt{v_r^2+v_{\phi}^2}$ (black), derived from the non-circular motions above.   Thin lines show the uncertainty measured from the errors above.  
The gas velocity dispersion profile $\sigma$--calculated as the azimuthal average of the line-of-sight dispersion corrected for unresolved bulk motions (i.e. \citealt{petyPAWS})--is shown for comparison (dotted).  
\label{fig:stream}}
\end{figure}

If our profile of the present-day torques in M51's inner disk is correct, then
we expect the signatures of the gas motions that occur in
response to the torques to be evident in the velocity field
derived from the CO data. In Appendix \ref{sec:appendixStream}, we use our mass-based model of the circular motion in M51 (Appendix \ref{sec:appxRot}) to isolate the contribution of non-circular motion to the line-of-sight velocity field of the CO emission.  
Then we decompose this non-circular motion into its radial and azimuthal components $v_r$
and $v_{\phi}$.  

Figure \ref{fig:stream} shows radial profiles of both reconstructed velocity components, which are calculated using CO intensity as a weighting factor in binned azimuthal averages to reveal the velocities characteristic of the (CO bright) spiral arms.  
Profiles with inverse-intensity weighting, which highlight the inter-arm streaming velocities, are qualitatively similar but show velocities $\approx$ 5-10 km s$^{-1}$ lower than the profiles in Figure \ref{fig:stream} (see Appendix \ref{sec:appendixStream}).  

Our reconstructed radial velocities present us with a picture of radial gas flow that is qualitatively similar to that implied by the present-day torques.  
We once again retrieve the pattern of radial inflow and (weak) outflow that mark the locations of at least two distinct CR, at nearly the same radii as found with the torque map (Figure \ref{fig:stream}).  
Radial velocities go to zero and switch sign near 25\arcsec ~and again between the range 60\arcsec$<$$R$$<$80\arcsec ~as predicted by the torques.  Meanwhile, the azimuthal velocities exhibit similar zero-crossings as well as a pattern of positive and negative values that reflect a complex response to the spiral arm; depending on location relative to the spiral, gas at some points is stationary in azimuth or even flows opposite to the direction of rotation.  
There is some dependence on the adopted rotation curve (and hence stellar M/L), which shifts the azimuthal velocity horizontally in Figure \ref{fig:stream}.  
But, as described below, this choice does not greatly alter the measured magnitude of the streaming motions.  

Almost independent of the particular rotation curve (or adopted stellar M/L; see Appendix \ref{sec:appendixStream}), the regions of negative torque identified in Figure \ref{fig:torqsfe} map to regions of the highest non-circular velocities in the arm.   
The magnitude of streaming motions $\vert v_S\vert=\sqrt{v_r^2+v_{\phi}^2}$ is shown in the bottom panel of Figure \ref{fig:stream} for each of our three possible rotation curves, i.e. assuming the nominal M/L=0.45$\pm$0.15 (chosen as described in Appendix \ref{sec:appendixStream}).   
Although we do not expect negative torques to coincide with the largest streaming motions as a rule, the link between $v_r$ and $v_S$ is expected, according to the predicted relation between radial and azimuthal velocities in response to a spiral potential perturbation (i.e. \citealt{bt}): for an $m$-armed spiral rotating at pattern speed $\Omega_p$ in a galaxy with a flat rotation curve (zero background shear) $v_\phi$=$v_r$$2\Omega/m(\Omega-\Omega_p)$ (with $\Omega$ the disk angular velocity), and so the total magnitude of streaming motions will follow $v_r$, even at CR since $v_r$$\propto m(\Omega-\Omega_p)$.  

\subsection{A Link Between Gas Flows and Star Formation Efficiency in M51's inner disk }

A number of previous studies have shown that the pattern of M51's star formation exhibits a strong radial dependence (see \citealt{schinner2013} and references therein).  
While star formation falls on the convex side of the spiral very clearly between 60\arcsec$\lesssim$$R$$\lesssim$80\arcsec (2$\lesssim$$R$$\lesssim$3 kpc), the young stars and gas coincide in the outer material arms.  
Meanwhile, there is little, if any, evidence of star formation in the central 20\arcsec ~(in the nuclear bar zone; but see below) and between 35\arcsec$\lesssim$$R$$\lesssim$60\arcsec ~in either H$\alpha$ or 24 $\mu m$ emission.\footnote{Both star formation tracers show a strong deficit in the zone 35\arcsec$\lesssim$$R$$\lesssim$60\arcsec, in particular, ruling out the extinction of optical photons as the source of the radial variation in the inferred SFR.}  
The radial dependence in star formation activity in M51 causes the measured star formation rate per unit of gas, or star formation efficiency SFE=$\Sigma_{SFR}/\Sigma_{gas}$ (i.e. \citealt{leroy08}), and hence gas depletion time $\tau_{dep}$=SFE$^{-1}$, to vary by as much as a factor of 6 throughout the inner 9 kpc.  
This same radial variation is evident in the SFE profile calculated by \citet{leroy}.  \citet{knapen} identified this same pattern and loosely interpreted this as a dependence of the SFE on location relative to spiral arm resonances.  

In Figure \ref{fig:torqsfe}, we compare the molecular gas depletion time $\tau_{dep}$ to the profile of present-day gravitational torques.  
To calculate $\tau_{dep}$, we divide the radial profile of the molecular gas surface density $\Sigma_{H2}$ by the star formation rate surface density $\Sigma_{SFR}$ profile.\footnote{The gas depletion time defined in this way is the time to deplete the molecular material, rather than the total gas, although the two timescales should be similar; the atomic gas surface density is a factor of 10 lower than the molecular gas surface density across the PAWS field of view (\citealt{schuster}; \citealt{leroy})}  
By using azimuthal averages of $\Sigma_{SFR}$ and $\Sigma_{H_2}$ we minimize the unavoidable bias introduced by the spatial offset between gas and young stars predicted by density wave theory--and well-known to appear along the spiral arms in M51.  

It is clear from Figure \ref{fig:torqsfe} that there is a strong link -- shown as an inverse correlation -- between $\tau_{dep}$ and the gravitational torques, underlining a strong dependence of star formation on dynamical environment. 
Radially inflowing gas appears to be less efficiently forming stars.  
More generally, gas sitting near CR forms stars more efficiently than gas that is away from CR and in motion relative to the background potential.  
We expect this to apply in other systems where radial outflow motions could potentially coincide with low star formation efficiency.  

In contrast to expectations for a universal gas depletion time (i.e. \citealt{bigiel} and \citealt{leroy}), it therefore appears that the $\tau_{dep}$ in M51 shows a genuine, and non-monotonic, radial dependence.  
This was previously demonstrated by \citet{schuster} out to 12 kpc, although with a slightly different radial trend and using a different star formation tracer.  In  particular, they find a depression in the ratio $\Sigma_{H2}$/$\Sigma_{SFR}$ in the zone of the nuclear bar.  However, several factors complicate reliable estimation of $\Sigma_{SFR}$, and thus $\tau_{dep}$, in this area, in particular.   The diffuse FIR contribution from dust heating by an older population of bulge stars, for one, may lead to overestimation of the SFR.  Age and/or metallicity gradients in the underlying old stellar population also lead to uncertainties in the stellar continuum subtraction of narrow-band H$\alpha$ imaging.  
 
In this work, we do not find the same rise in $\Sigma_{SFR}$ toward the center as \citet{schuster}, possibly because any diffuse component present in the 24 $\mu m$ image we use is surreptitiously minimized in combination with the continuum-subtracted H$\alpha$, which shows a central deficit \citep{schinner2013}. 
In light of these uncertainties in $\Sigma_{SFR}$, we consider our estimates for $\Sigma_{SFR}$ and $\tau_{dep}$ to be least reliable in the center.  
We note, though, that an increase in $\tau_{dep}$ toward the center, such as we find, might be expected given the high shear characteristic of these radii (see Figure \ref{fig:shearplot}).  

Throughout the remainder of the paper we focus primarily on the the spiral arms, which show persistent variation in $\tau_{dep}$ that is more arguably real.  
The gap in star formation between 35\arcsec$\lesssim$$R$$\lesssim$60\arcsec ~is evident in multiple tracers, in contrast with the high $\Sigma_{gas}$ in this zone.  
This contrast is particularly obvious at the high resolution of the PAWS observations \citep{schinner2013}. 
But even at lower resolution, it is quantitatively clear that the star formation rate is lower than expected for the relatively high gas column; fewer stars are formed than expected for a global star formation law, or universal $\tau_{dep}$.  
Furthermore, the pattern in $\tau_{dep}$ can not be attributed to variation in $X_{CO}$.  There is little to no indication that $X_{CO}$ varies across the PAWS field of view (indeed, M51 has a negligible metallicity gradient), as recently discussed in detail by \citealt{colombo2013}).   
Instead, we propose that the radial dependence in $\tau_{dep}$ reflects the influence of dynamics on cloud properties (see Section \ref{sec:dynamicalPquant}), following the large-scale organization of the ISM by the non-axisymmetric structure present in the disk.  

\section{GMC Evolution in M51: The Importance of Dynamical Pressure on Cloud Stability and star formation}\label{sec:dp}

\subsection{Dynamical Pressure: Qualitative Remarks \label{sec:dynamicalPqual}}

\begin{figure}[t]
\includegraphics[width=1\linewidth]{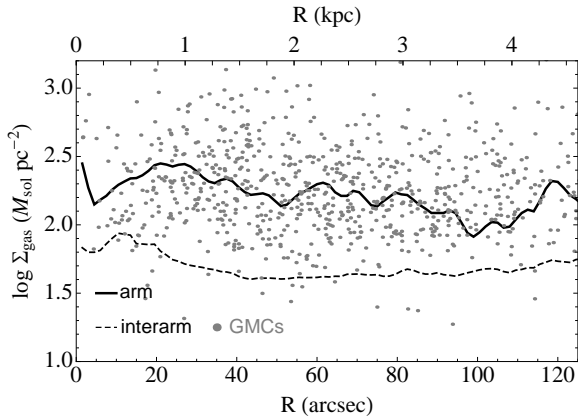}\\
\caption{Molecular gas surface densities throughout the PAWS field.  Catalogued GMCs (see \citealt{colombo2013}) are shown as gray dots, together with the average surface density in the arm region defined as in \citet{colombo2013} (black thick curve) and the average inter-arm surface density, calculated via inverse-intensity weighting (dashed thin curve). }
\label{fig:surfs}
\end{figure}

In the previous Section, we showed that there is a strong link between
zones of strong gas flow (which, in M51, correspond to regions of radial inflow and negative torque) and regions where molecular gas appears to be
inefficient at forming stars. This is especially striking in the 
inner spiral arms spanning 35\arcsec$\lesssim$$R$$\lesssim$60\arcsec, where the molecular gas reaches relatively high
surface densities but is effectively devoid of high-mass star
formation activity, in stark contrast to models that propose that the
surface density of SFR and gas are well-correlated.  

In particular, Figure \ref{fig:stream} shows that the molecular gas exhibits a complex
response to M51's underlying non-axisymmetric potential.  The rates at
which gas flows along and across the arms depend on position along the
arm, and these rates vary in magnitude relative to one another
depending on distance from corotation. The length of time spent in the
potential well of the arm depends on the perpendicular component of
the velocity $v_\perp$, which is well-approximated by the radial streaming velocity $v_r$ given the
tightness of the spiral.  How far gas travels along the arm before
exiting on the downstream side depends on the azimuthal streaming
velocity $v_\phi$.  The smaller this rate, the more quickly gas moves
in to and out of the spiral, leaving little time for clouds to acquire
mass (i.e. through collisions/agglomeration in the spiral arm), become
gravitational unstable and form stars.  At corotation, both velocities
are zero and gas is stationary in the rotating frame.

We can begin to appreciate the effect of such motion on clouds by
considering the ratio $v_S^2$/$\sigma^2$, which is a ratio of
timescales: the velocity dispersion $\sigma$ sets the 'equalization'
time, or the timescale for clouds to reach pressure equilibrium with
their surroundings, and the streaming velocity $v_S$ sets a travel, or relocation, time.  In
this case, $v_S^2$/4$\sigma^2$ is a measure of the square of the ratio
between the time it takes to pressure-equalize and the time to cross
two cloud lengths.\footnote{We adopt the two-cloud length crossing
  time for this particular ratio for consistency with the formalism in
  $\S$\ref{sec:dynamicalPquant}} With the introduction of non-circular motions
$v$$\gtrsim$$\sigma$, clouds equalize less quickly simply because they
do not remain in any one environment for long enough.

This fact is underlined in the bottom panel of Figure
\ref{fig:spiralframe}, which shows that the ratio of equalization to
relocation timescales is large everywhere except near CR where, by
definition, gas is motionless in the rotating frame.  We therefore
interpret the close (if qualitative) relation between $\tau_{dep}$ and
$v_S^2$/$\sigma^2$ to suggest that moving clouds have insufficient
time to collapse and form many stars. This can be viewed as an
effective stabilization, quantifiable in terms of dynamical pressure
$P_{dyn}=\rho v_S^2$.  
As described in the next section, dynamical pressure reduces the internal pressure of the streaming medium in
which clouds are embedded.  This reduces the surface pressure on
clouds, countering the gravity of the cloud.

According to the timescale arguments above, the dynamical pressure does not `dissipate' in a cloud-crossing time, implying that dynamical pressure can not be ignored.  
Of course, whether or not clouds are sensitive to changes in surface pressure depends on whether clouds are dynamically decoupled from their surroundings.  This depends on the internal pressure of the cloud in relation to the external pressure.  
Observations of clouds at high density contrast to the surrounding medium are usually taken to imply an extreme pressure imbalance, and thus little sensitivity to environment (e.g. \citealt{krumholz12} and references therein).  
However, 
Hughes et al. (2012) have recently shown that the internal and external pressures of clouds in M33, the LMC and M51 (among others) track each other, even if they may not be equal, similar to the clouds in M64 studied by \citet{RosolowskyBlitz05}. 

We thus have reason to expect that pressure plays an especially important role in M51, not least because the high molecular to atomic gas ratio signifies a very high midplane pressure (e.g. \citealt{BlitzRosolowsky04}).  
This was discussed by \citet{hitschfeld} who find that the hydrostatic pressure is tightly correlated with the fraction of molecular to atomic material, indicating that this physical parameter determines the formation of molecular gas from  the atomic phase.  

As shown in Figure \ref{fig:surfs} the surface density in the ambient ISM is comparable to, and in some places exceeds, the surface density of individual GMCs.  
This implies that the internal and external cloud pressures are similar, and so we expect that clouds embedded in this medium will be to some degree pressure supported (see, e.g., \citealt{krumholz}).   
A large fraction of  clouds from across the PAWS field do, in fact, appear to be supported by pressure according to their measured virial parameters $\alpha$$\gtrsim$1  \citep{colombo2013}.  Clouds from the inner 3 kpc where the surface densities in the arm lie well above $\Sigma_{H_2}$=100 $M_\odot$ pc$^{-2}$ preferentially exhibit even higher $\alpha$$\gtrsim$2 than clouds from elsewhere in the disk.  
These clouds, which are comparable to clouds under starburst conditions, will therefore be keenly aware of changes in surface pressure.  

Even in lower surface density environments, changes in surface pressure will effect cloud equilibrium, making the difference between stability and collapse.  
This may be the case in other spiral galaxies with lower molecular gas fractions than M51 since, 
even in these galaxies, {\it arm} surface densities will often lie near $\Sigma_{GMC}$, even if the average surface density does not, preventing clouds from completely decoupling from their surroundings.  For example, if we write the surface density of a gas disk hosting a two-armed spiral as
\begin{equation}
\Sigma(R,\phi)=\Sigma_0(R)\cos(\phi)^{2p}
\end{equation}
where higher (even) $2p$ describes a thinner spiral,
then the average gas surface density 
\begin{equation}
<\Sigma(R)>=\frac{1}{2\pi}\int_0^\infty\Sigma(R,\phi)d\phi=\Sigma_0(R)\frac{(2p)!}{4^p p!p!}
\end{equation}
will be $\sim$0.25-0.5 times the peak surface density in the arm.  
According to the surface densities profiled by Leroy et al. (2008) we expect that, within the molecule-dominated zones in spiral galaxies, the gaseous spiral arms will supply non-negligible pressure support for GMCs. 

\subsection{Dynamical Pressure: A Model Based on the Bernoulli Principle\label{sec:dynamicalPquant}}

In the previous section we established a qualitative correlation between the gas depletion time and dynamical pressure, which suggests that changes in the cloud surface pressure induced by gas kinematics influence cloud stability, and therefore the ability of a cloud to form stars.  
To describe how these changes emerge 
we next develop a simple model based on the Bernoulli principle, which equates an increase in gas velocity with reduced pressure.  

Specifically, we work under the assumption of isentropic flow of a compressible isothermal ideal gas and relate gas properties along the streamline between two zones using the compressible Bernoulli equation.  
This is a simplification and ignores the fact that gas in reality undergoes shocks and other forms of energy loss; gas in perturbed disks moving at non-zero velocity relative to the background rotating frame will exhibit transonic speeds (Mach number $M$$>$1).  
The shock can be weakened by strong gas self-gravity (e.g. \citealt{kko}; \citealt{wada}) and we expect this to be true in M51.  
But in the frame of the shock, entropy is still not conserved and the values for the quantities assumed in the expressions in the following section will depart from the values assumed (the less so the weaker the shock).  

In particular, under the assumption that the dominant source of pressure is turbulent in nature so that $P$=$\rho  \sigma_{turb}^2$, we assume that the internal turbulent motion of the gas is uniform throughout the disk and take the average velocity dispersion $\sigma_{turb}$$\approx$10 kms$^{-1}$ observed in M51 \citep{colombo2013v} as representative of this motion.  
This value takes into account the contribution of unresolved bulk motions to the observed linewidth, which we subtract (following the formalism adopted by \citealt{petyPAWS}).  
Removing these motions yields a much more uniform velocity dispersion, although there is still a factor of 2 difference in the turbulent motions between the arm and interarm, likely because of the shock.  
Since this factor of 2 contrast is smaller than the density contrast between arm and interarm (typically 5-10), we ignore it.  We can more closely approximate reality by letting the density depend on pressure, i.e. by assuming compressibility.   

Only with a better observational handle on energy losses and shocks in gas (which is minimal at present) will it be possible to invoke an expression equivalent to Bernoulli but with the change in entropy included.  
This is something for which numerical calculation is better equipped and so we proceed with our simple calculation below with the aim of providing at least a coarse description of the phenomenon.

\subsubsection{The stable GMC mass threshold\label{sec:stablemass}}

In this section we compare the maximum cloud mass stable against collapse in two different flows, i.e. in the arm and inter-arm regions.  
Gas moving at high relative velocity, tunneled along and through the arms, will result in a reduced external surface pressure for clouds forming in the arm.  
This then increases the maximum mass stable against collapse, as related by expression for the Bonnor-Ebert mass $M_{BE}$$\propto P^{-1/2}$ calculated from pressure, rather than virial, equilibrium, i.e. including a (thermal) surface pressure term. \footnote{Note that the Bonnor-Ebert mass is equivalent to the Jeans mass ignoring any external or dynamical pressure.}$^,$\footnote{Even if the surface term were of kinetic origin (rather than thermal, as assumed in the expression for the Bonnor-Ebert mass) we would expect an analogous change in stable mass with a change in surface pressure, such as considered by \citet{paredes} }  
As a result of the dynamical pressure 1) any collapse-unstable clouds that form will be of higher mass than virialized clouds and 2) fewer clouds will be collapse-unstable, leading to lower SFE.  

We can estimate the fractional increase in the Bonnor-Ebert mass as the result of dynamical pressure by equating the energy densities of gas in two zones with and without strong streaming motions (e.g. in the arm and interarm) in the non-intertial (rotating) reference frame.  
For simplicity, we use the compressible Bernoulli equation to relate gas properties along the streamline between the two zones.  
For pressure $P$ dominated by turbulent motion with dispersion $\sigma$ (i.e. $P_i$=$\rho_i  \sigma^2$; note that this ignores an enhancement in cloud internal turbulence during passage through the arm), the pressure ratio of gas in zone 1 and zone 2 can be expressed as
\begin{equation}
\ln \frac{P_1}{P_2}=\frac{v_2^2}{2 \sigma^2}-\frac{v_1^2}{2 \sigma^2}+\frac{\sqrt{2\pi}G}{\sigma^2} (\Sigma_2h_2-\Sigma_1 h_1)
\label{eq:dyn1}
\end{equation}
with $h_i$ the gas scale height and $\Sigma_i$ the gas surface density in the two zones.  
The first two terms on the right are the dynamical pressures, while the third represents the change in the hydrostatic pressure from zone to zone.   Omitted terms include the stellar density factor in the hydrostatic pressure and other energy losses/gains.   

As we are near enough to hydrostatic equilibrium so that $\Sigma_1/\Sigma_2$$\approx$$h_2/h_1$, the third term on the right is negligible and the first two terms dominate.  

Since fractional increase in the stable (Bonnor-Ebert) cloud mass is
\begin{equation}
\frac{d M}{M_{1}}=-\frac{d P}{2P_1}\nonumber
\end{equation}
then 
\begin{equation}
\ln \frac{M_2}{M_{1}}\approx\frac{v_2^2-v_1^2}{4 \sigma^2}\label{eq:full}
\end{equation}

From this point we can see that streaming motions raise the cloud stable mass above the virial mass by \begin{equation}
\log \frac{M}{M_{vir}}\approx\frac{v_S^2}{4 \sigma^2}\frac{1}{\ln{10}},
\end{equation}
where the non-circular streaming in zone 2 is given by $v_S$ and zone 1 is identical to zone 2 except now its clouds are free of streaming (and dynamical pressure) and thus assumed to be virialized.  

Within gas rotating in the plane of the galaxy, the stable mass is raised for all clouds undergoing non-circular motion.  
The stable mass for clouds moving at streaming velocity $v_{S}$$\sim$15-20 km s$^{-1}$, typical of spiral arm streaming in M51, will be at least 1.5-2 times higher than for {\it virialized} clouds with similar surface densities and line widths $\sigma$$\lesssim$10 km/s.  
As a result of the reduced surface pressure, 
collapse-unstable clouds that go on to form stars will be on average 1.5-2 times more massive than virialized clouds.    

The difference from arm to interarm cloud masses should be similar, given the weaker streaming in the interarm.   
Conservatively allowing a factor of 1.5 difference in $v_{S}$, then we expect bound cloud masses to be on average 0.5 dex higher in the arm than in the interarm, very nearly what is measured from the cloud-decomposed PAWS CO(1-0) emission (see \citealt{colombo2013}; evident from the horizontal offset in the mass spectra for clouds in the two zones).  
Again, this ignores real variation in the gas dispersion from arm to interarm, but we note that the change in measured cloud line-width between these two environments in M51 is only modest (\citealt{colombo2013})
\begin{figure*}[t]
\begin{tabular}{ll}
\begin{minipage}[b]{.5\linewidth}
\hspace*{-.1in}\includegraphics[width=\linewidth]{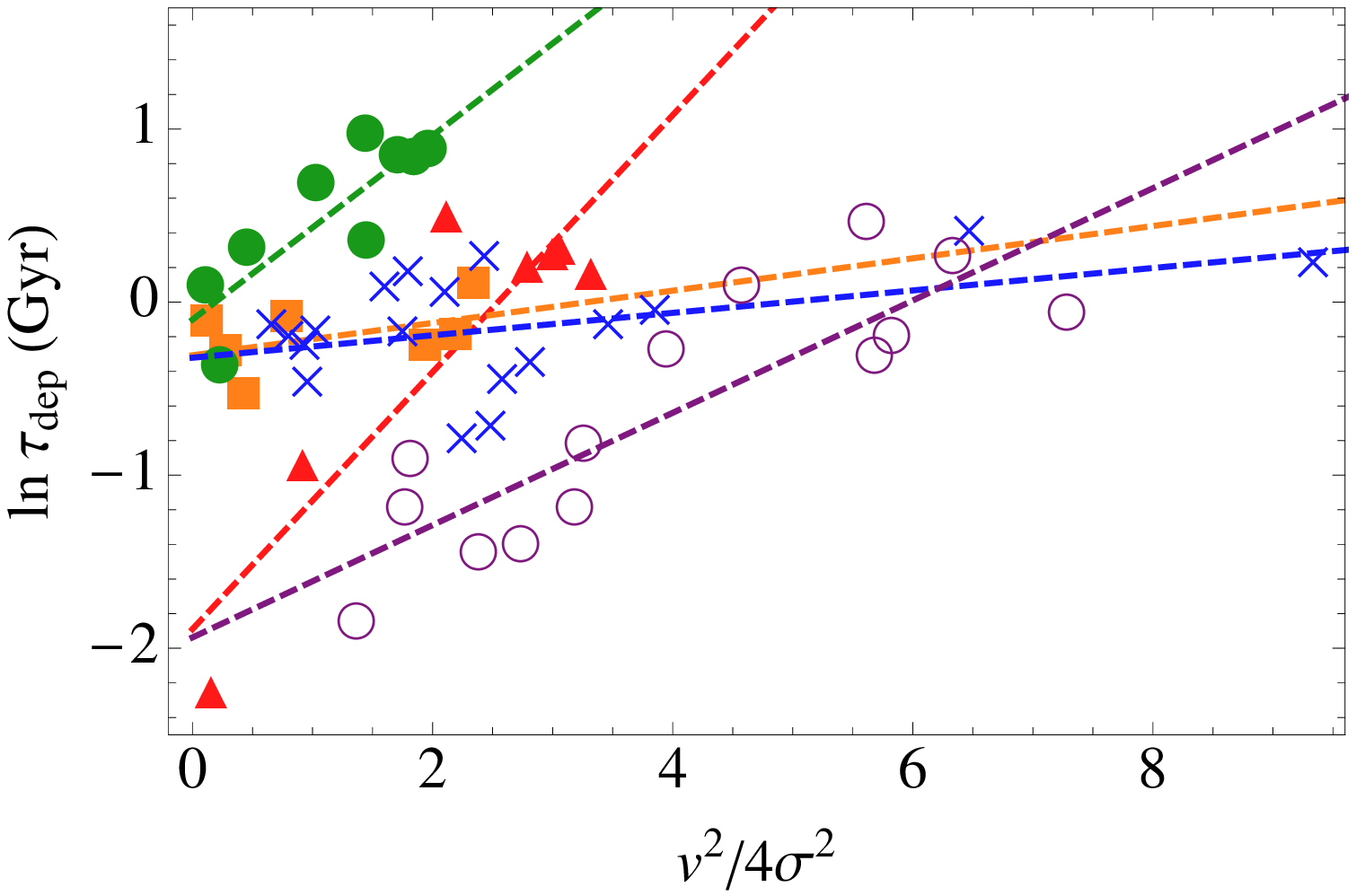}
\vspace*{.15in}
\end{minipage}&\hspace*{-.1in}\includegraphics[width=.5\linewidth]{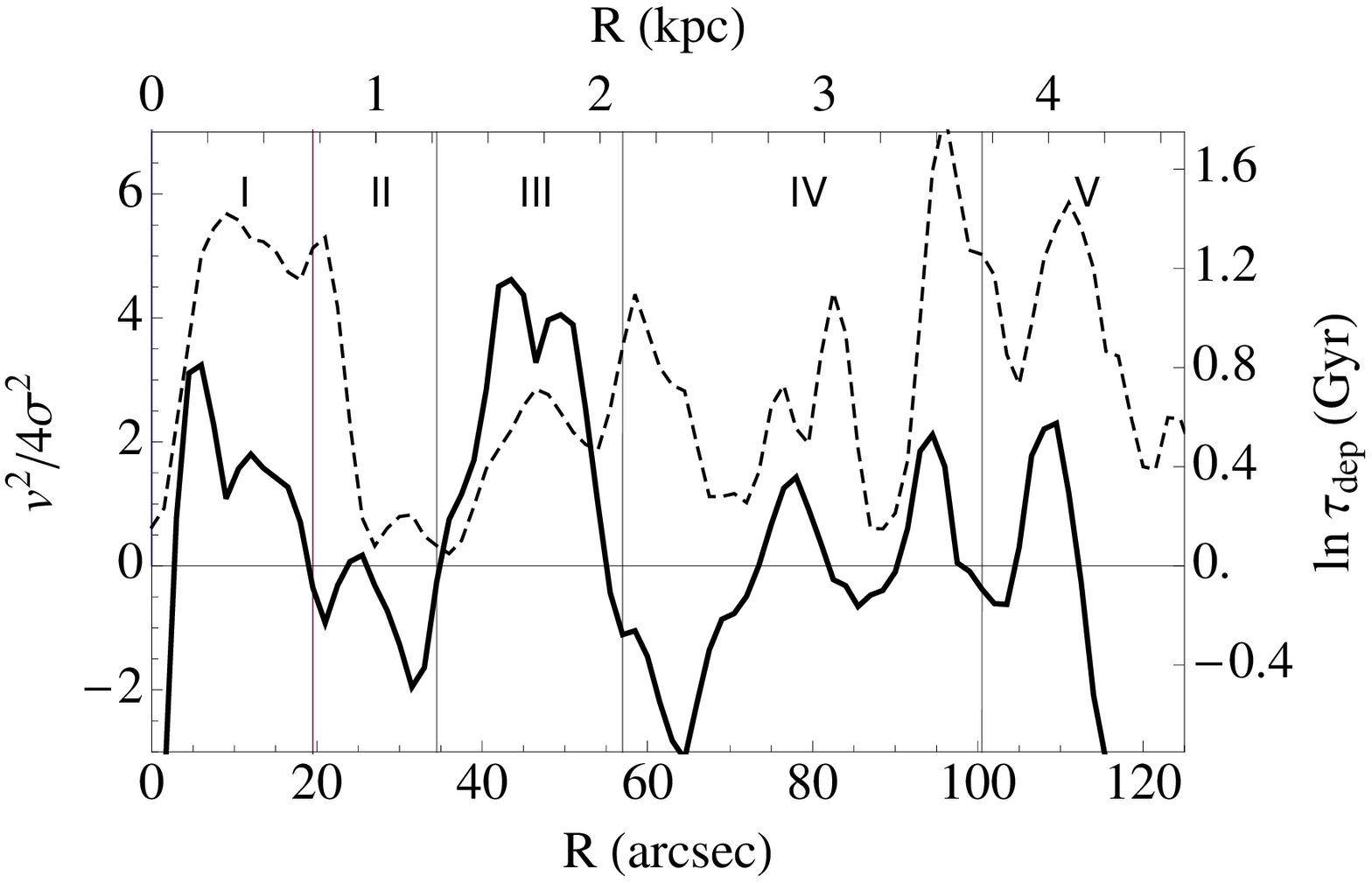}
\vspace*{-.25in}
\end{tabular}
\begin{tabular}{c}
\includegraphics[width=1\linewidth]{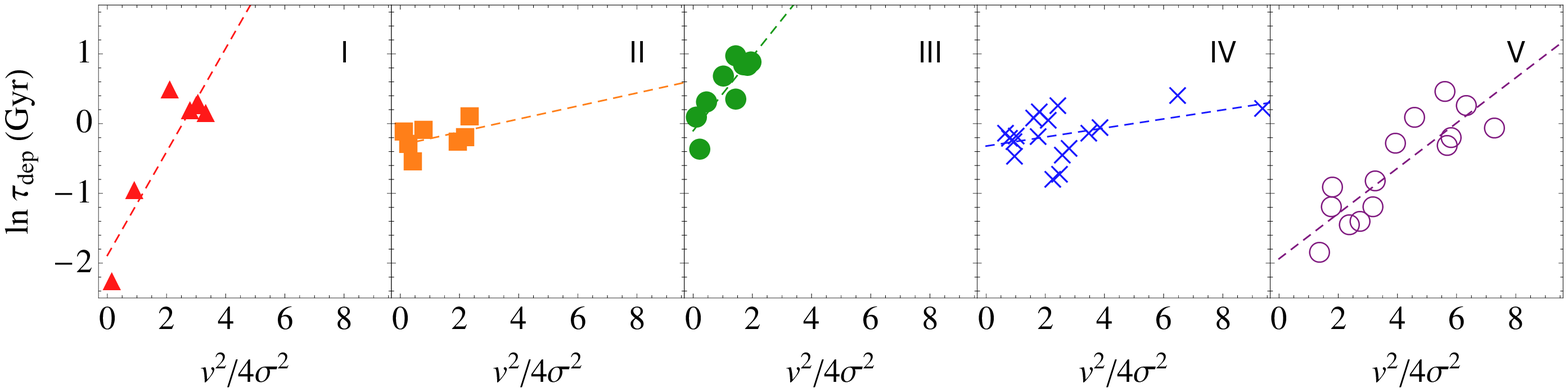}
\end{tabular}
\begin{centering}
\caption{(Top Left) Scatter plot of $\ln \tau_{dep}$ vs. $v_S^2$/4$\sigma^2$ extracted in 2.4\arcsec ~bins from the curves at right.  The symbols and color scale varies as at right, according to dynamical environment.  
Best-fit lines for each environment are overlayed. (Top Right) Profiles of $\ln \tau_{dep}$ (solid black) and $v_S^2$/4$\sigma^2$ (dashed black) throughout M51.  The measurements of $v_S$ and $\sigma$ used here are shown in the bottom panel of Figure \ref{fig:stream}.  Errors have been omitted for clarity.  Dynamical environments defined in $\S$ \ref{sec:env} (and described by \citet{colombo2013}) are designated by color overlays: nuclear bar (red), molecular ring (orange), spiral inside CR (green), spiral outside CR (blue), material pattern (purple).  
(Bottom) Scatter plots of $\ln \tau_{dep}$ vs. $v_S^2$/4$\sigma^2$ separated by environment.  \\}
\label{fig:velsfe}
\end{centering}
\end{figure*}
\begin{table}
\caption{Environment-dependent correlations\label{tab:fits}}
\begin{center}
\begin{tabular}{cccc}
\tableline\tableline
environment&$\gamma$&$\tau_{dep,0}$&$\chi^2$\\
\tableline
I& -1.75$\pm$0.16& 0.15$\pm$0.06&5.30\\
II& -1.1$\pm$0.18& 0.73$\pm$0.21&0.88\\
III&-1.54$\pm$0.24&0.89$\pm$0.28&1.98\\
IV& -1.07$\pm$0.05& 0.72$\pm$0.13&6.85\\
V& -1.33$\pm$0.08& 0.14$\pm$0.04&7.43\\
\tableline
environment&$\gamma$&$\tau_{dep,0}$&$\chi^2$\\
\tableline
I& -1.82$\pm$0.33& 0.52$\pm$0.12&21.81\\
II& -1.08$\pm$0.19& 0.79$\pm$0.15&0.96\\
III&-1.48$\pm$0.22&1.54$\pm$0.25&2.16\\
IV& -1.07$\pm$0.05& 0.75$\pm$0.11&6.6\\
V& -1.15$\pm$0.04& 0.43$\pm$0.06&13.35\\
\tableline
environment&$\gamma$&$\tau_{dep,0}$&$\chi^2$\\
\tableline
I&$\hdots$& 0.11$\pm$0.02&9.25\\
II&$\hdots$& 0.52$\pm$0.06&26.74\\
III&$\hdots$&0.68$\pm$0.11&14.03\\
IV&$\hdots$& 0.14$\pm$0.02&112.75\\
V&$\hdots$& 0.04$\pm$0.01&108.7\\
\tableline
\end{tabular}
\end{center}
\tablecomments{\small Estimates of the cloud mass spectrum index $\gamma$ and fiducial gas depletion time $\tau_{dep,0}$ measured from the slope and intercept, respectively, of the best-fitting straight line in each of the 5 distinct dynamical environments indicated in Figure \ref{fig:velsfe}.  The $\chi^2$ value of the fit is given in the right-most column.  
The five environments are, in ascending order, nuclear bar, molecular ring, spiral inside CR, spiral outside CR, and outer material pattern.   The top third of the table lists the results of the fits shown in Fig. \ref{fig:velsfe}.  
The middle third lists the results of fits to similar points, but that take into account inter-arm streaming, and the bottom third lists the $\tau_{dep,0}$ measured assuming a fixed slope corresponding to $\gamma$=-1.72, the average index measured via direct fitting of the cloud mass spectra in these same environments (Hughes et al. 2012).  }
\end{table}
\begin{figure}[t]
\begin{centering}
\includegraphics[width=1.12\linewidth]{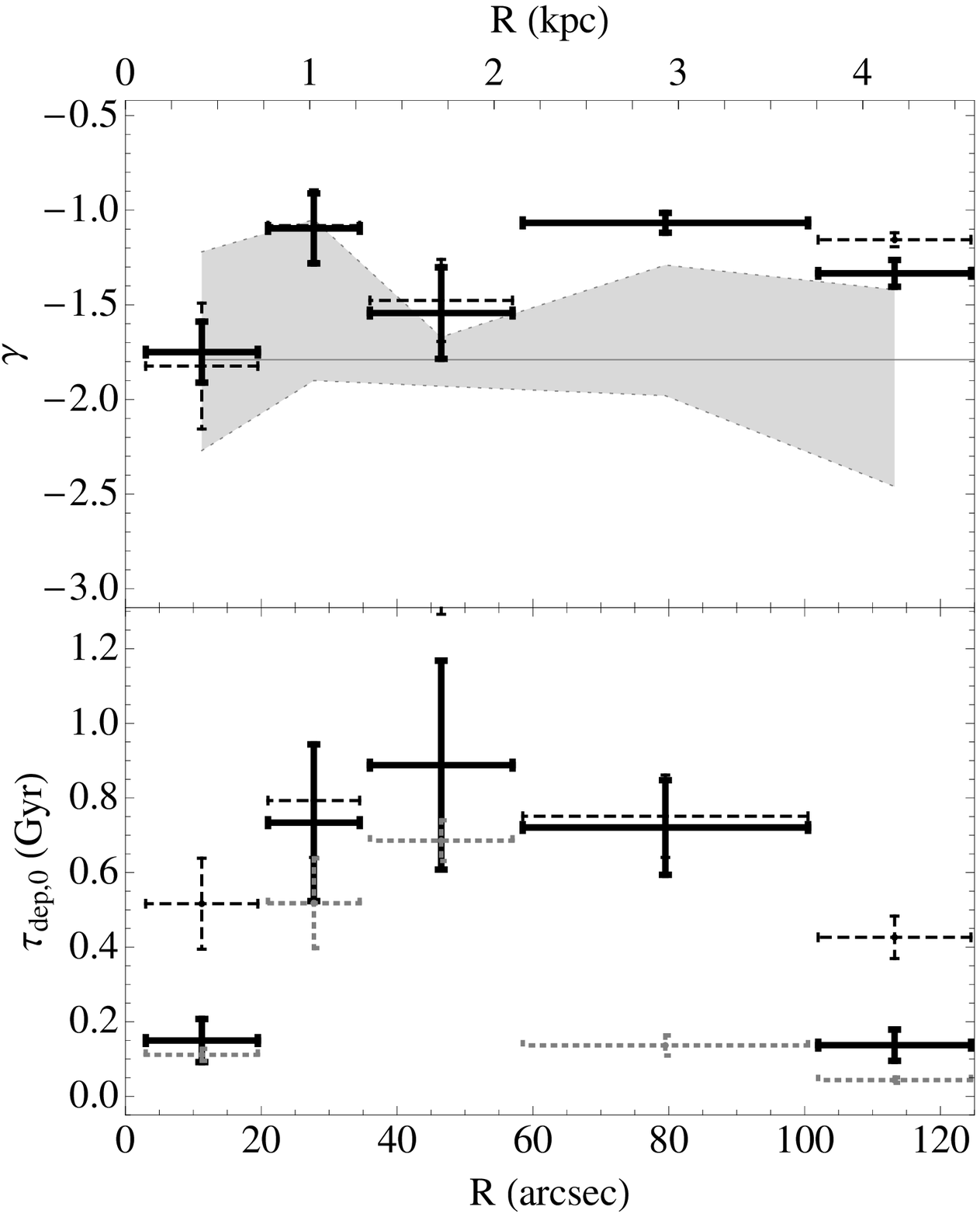}
\caption{(Top) Cloud mass spectrum power-law index from within the series of M51 dynamical environments designated in Figure \ref{fig:velsfe}.  
The center of each environment is taken as the radial position of each measurement.  
Indices estimated from two independent approaches are shown: 1) the range spanned by the minimum and maximum indices measured from direct fits to the cloud mass spectra in each environment is shown in light gray (\citealt{colombo2013}; \citealt{hughes}).  
The gray horizontal line marks the average $\gamma$=-1.72.  2) The indices implied by the best-fit slopes in Figure \ref{fig:velsfe} in each environment are shown in solid black with errors given by the statistical error in each measured slope.  Best-fit slopes accounting for inter-arm streaming velocities are shown in thinner dashed line. 
(Bottom) Fiducial gas depletion time $\tau_{dep,0}$ in the same series of M51 dynamical environments measured from the best-fit intercepts in the right panel of Figure \ref{fig:velsfe} (solid black), the best-fit intercepts when interarm streaming velocities are included (thin dashed black) and the best-fitting $\tau_{dep,0}$ with fixed $\gamma$=-1.72 (dotted gray).
}
\label{fig:slopes_ints}
\end{centering}
\end{figure}

\subsubsection{Impact on $\tau_{dep}$\label{sec:impactSFE}}

In the previous section we showed that, as a consequence of strong non-circular streaming motions, the collapse-unstable clouds that go on to form stars will be on average 1.5-2 times more massive than virialized clouds.  
This suggests that star-forming clouds in the arm (where streaming is high) preferentially host massive star formation.  But at the same time, any clouds in the arm below $\sim$2 times the virial mass will not collapse and form stars.  
This naturally translates in to a decrease in the star formation efficiency, considering the change to the fraction of the total gas reservoir in the form of collapse-unstable clouds per free fall time.   

Specifically, assuming that the process (or processes) responsible for the build-up of GMCs results in a mass spectrum \begin{equation}
N(>M)\propto\left(\frac{M}{M_{max}}\right)^{\gamma+1} \label{eq:massspec} 
\end{equation}
then from the mass in (unstable) clouds above the stable threshold 
\begin{equation}
M(>M_{stable})=\int_{M_{stable}}^{M_{max}}\frac{dN(M)}{dM}MdM\\
\end{equation}
we see that the mass fraction of such clouds $F=M(>M_{stable})/M_{total}$ in a region of total mass $M_{total}$ will change as
\begin{equation}
d F=(1+\gamma)\left(\frac{M}{M_{max}}\right)^{\gamma}\frac{d M}{M_{total}}
\end{equation}
with a change to the stable mass by an amount $dM$.  

Now defining the average unstable cloud mass $<$$M_c$$>$=$M(>M_{stable})/N$, then 
\begin{equation}
\frac{d F}{F}= (1+\gamma)\frac{d M}{<M_c>}.
\end{equation}

We can cast this in terms of the SFE by recognizing that the collapse unstable clouds--those above the stable threshold--are the clouds responsible for star formation.  
Specifically, we write SFE=$F\beta$ where $\beta$ is the fraction of a cloud that is converted in to stars per free fall time (typically less than 5\%; e.g. \citealt{krumholzmckee}; \citealt{murray2011}).  In this case, the fractional change to $\tau_{dep}$, the inverse of SFE, is given by $d \tau_{dep}/\tau_{dep}=-d F/F$
whereby
\begin{eqnarray}
\ln\frac{\tau_{dep}}{\tau_{dep,0}}&\approx& -(1+\gamma)\ln\frac{M}{<M_c>}\approx -(1+\gamma)\ln\frac{M}{M_{vir}}\nonumber\\
&=&-(1+\gamma)\frac{v_S^2}{4\sigma^2}\label{eq:prediction}
\end{eqnarray}
using that $<$$M_c$$>$$\approx$$M_{vir}$ in the absence of streaming and where constant of integration $\tau_{dep,0}$ is the unperturbed gas depletion time.  By design, we take $\tau_{dep,0}$ to be the depletion time in the case of virialized clouds.

\subsubsection{Comparing Model Predictions to Observations\label{sec:emp_corr}}

According to eq. (\ref{eq:prediction}), the gas depletion time, or inverse SFE, along the arm should proportionately follow the raised stable mass threshold in the presence of spiral streaming as measured by $v_S^2$/4$\sigma^2$.  
These two quantities are plotted as a function of radius in the left panel of Figure \ref{fig:velsfe}, using our measurements of the magnitude of streaming motions shown in the bottom panel of Figure \ref{fig:stream} and the observed velocity dispersion $\sigma$,\footnote{That is, we assume $\sigma$ describes the overall state of the gas (i.e. the combined thermal and kinetic pressure on the gas), including the contribution from turbulent motions.  
Note that, independent of streaming, gas with higher $\sigma$ is expected to contain clouds with higher virial masses.} which is found to be relatively uniform across the disk, near 10 km s$^{-1}$ (see Figure \ref{fig:stream}).  
With our very simple prediction we are able to reproduce the observed pattern of star formation;
the radial variation in the $\tau_{dep}$ nearly echoes the radial dependence in the kinematic term $v_S^2/\sigma^2$ to within a radially varying scale factor -($\gamma$+1). 

The close match (albeit with scatter) is underlined on the right panel of Figure \ref{fig:velsfe}, where points fall along environment-unique lines with slopes -(1+$\gamma$), according to eq. \ref{eq:prediction}.  
Table \ref{tab:fits} lists the slopes and intercepts of the best-fitting linear relationship in each of the five environments indicated in Figure \ref{fig:velsfe} determined with the FITEXY routine \citep{press} assuming uncorrelated errors in both variables.  We adopt 0.2 dex uncertainty in the measured $\tau_{dep}$, following \cite{leroy08}, and the errors plotted in Figure \ref{fig:stream} for the uncertainty in $v_S^2/4\sigma^2$.  
Despite the presence of scatter, the fitted relations are robust to changes in the radial range over which each dynamical environment extends.  Increasing or decreasing the size of each by 10\% leads to typically less than 10\% variation in the fitted slope and intercept.   

The average index $\gamma$=-1.32$\pm$0.26 measured from the best-fit slopes is consistent with (if slightly lower than) the indices measured in the same set of environments by \citealt{hughes} via direct fitting of the cloud mass spectra ($<$$\gamma_{df}$$>$=-1.72$\pm$0.39 on average).   
A 2$\degree$ error in the adopted inclination angle can easily explain this difference (see $\S$ \ref{sec:corr_quality}).  

Moreover, the measured indices are remarkably compatible with the genuine, if modest, trend in the `direct-fit' $\gamma_{df}$ with dynamical environment.  
In the study of \citet{colombo2013} variations in mass spectrum shape between the center, arm and interarm environments is interpreted in terms of cloud formation, growth and destruction mechanisms.  
We retrieve the record of these processes across the environments sampled here with our simple expression (but see $\S$ \ref{sec:corr_quality}).  

Meanwhile, the best-fit intercepts of each environment-dependent relation correspond to a fiducial gas depletion time $\tau_{dep,0}$=0.5 Gyr, on average.  
This value represents the time to consume the molecular gas in the absence of spiral streaming and is notably shorter than the average depletion time across the PAWS field of view, $<$$\tau_{dep}$$>$=1.5Gyr.  
The latter value is consistent with molecular gas depletion times measured in the THINGS sample of nearby galaxies by Bigiel et al. (2011) and Leroy et al. (2008, 2012).   
In some environments the fiducial $\tau_{dep,0}$ appears to be even shorter than 0.5 Gyr (see Figure \ref{fig:slopes_ints}; but see $\S$ \ref{sec:corr_quality}).  
Later in $\S$ \ref{sec:discussion} we develop a picture that attributes departures from a fiducial, universal gas depletion time to motions in non-axisymmetric potentials.  

We emphasize that the trend traced out in the right panel of Figure \ref{fig:velsfe} does not arise with radial gradients in the kinematic term $v_S^2/4\sigma^2$ or $\tau_{dep}$ across the PAWS field.  
Variation appears on smaller (arcsecond) scales, and the agreement between predicted and observed $\tau_{dep}$ is genuine.  
This may be surprising, given that our prediction is oversimplified, not least because it implicitly assumes a fixed, global power-law mass spectrum.  
As revealed by \cite{colombo2013}, cloud mass spectra in M51 show marked deviations from power-law behavior and uniformity along the arms coincident with changes in the dynamical environment defined here.  
These differences appear to indicate real variations in the mechanisms behind GMC formation, but they may also reveal how dynamical pressure plays a role in shaping/modifying an underlying global power-law mass spectrum (and lognormal intensity pdf; see Hughes et al. 2012).   
Detailed modelling will be required to better understand and distinguish these scenarios.   

\subsubsection{Sources of Uncertainty: Sensitivity to Observables\label{sec:corr_quality}}

Several factors influence the tightness of the correlation between $\tau_{dep}$ and $v_{S}^2$.  In the previous section we demonstrated that the overall scatter is dominated by real environmental variation in $\gamma$ and $\tau_{dep,0}$.  
Within a given environment, the scatter that emerges tends to be on the order of 5-7\%, according to the errors on the measured slope and intercept of the best-fit linear relation.  
Both within and among environments, the correlation is subject to observational uncertainty,   
namely to the accuracy in the star formation and gas density tracers, the kinematic parameters assumed to deproject the line-of-sight velocities and the pitch angle adopted in the calculation of the non-circular streaming motions.  

The reported statistical errors in the fitted $\tau_{dep,0}$ and $\gamma$ primarily reflect uncertainty in the assumed pitch angle and rotation curve, which together define the errors in $v_{S}^2$ used as weights in the reduced-$\chi$$^2$ fit to the linear relation in eq. (\ref{eq:prediction}).  
In testing we found that pitch angle variation is responsible for the largest uncertainty in the decomposition of the line-of-sight velocity field: by comparison, the radial and azimuthal streaming components are surprisingly robust to changes in the kinematic parameters, i.e. to the major axis position angle and inclination.  
We estimate that PA variation of $\pm$5$\degree$ introduces 10\% uncertainty in the magnitude of the streaming motions and therefore up to 20\% change in the best-fit slope.   
Velocities extracted via solution in the spiral arm frame exhibit the expected $\sin{i}$ dependence.   
We expect inclination uncertainty, unlike pitch angle, to apply globally (under the assumption of an unwarped disk), resulting in an overall shift in the fitted $\tau_{dep,0}$ and $\gamma$.  
A slightly higher inclination of $i$=23$\degree$ would bring the average $\gamma$ in to perfect agreement the average index measured by Hughes et al. (2012). 

To construct our estimate of the star formation rate, we use a combination of H$\alpha$ and 24 $\mu m$ emission.  
This accounts for obscured and unobscured star formation but does not explicitly include a correction for the diffuse component of the 24 $\mu m$ emission (e.g. as explored by Leroy et al. 2011), which is thought to arise with dust heating by an underlying older population of stars.  
The central bulge region (but also the spiral arms) most likely incorporates some level of dust emission that does not trace young stars, in light of the enhanced old stellar surface densities in this zone (i.e. as demonstrated in the bulge region of M31 by Groves et al. 2012).   
Underestimation of $\tau_{dep}$ in the center as a result is likely responsible for the particularly steep power law index estimated for the nuclear bar environment, as well as the low $\tau_{dep,0}$ there.  

A low $\tau_{dep}$ in the center could alternatively arise from underestimation of the molecular gas surface density.  
Given the excellent sensitivity to low-level CO emission in the PAWS data set and little expected variation in $X_{CO}$ (\citealt{hughes}; \citealt{colombo2013}) the latter would most likely occur only if the CO fails to trace a warmer molecular phase, e.g., as more regularly observed in the centers of galaxies (Dumas et al.).   
The gas depletion time in the outermost material arm environment may also be artificially lowered due to the increased contribution from atomic gas component omitted at these radii (although this is still below 10\% at the edge of the PAWS field; see the profiles in Leroy et al. 2008).  

Finally, the quality of the correlation depends on the reliability of our assumptions.   We ignore any non-uniformity in the turbulent motions from arm to interarm as well as energy losses that occur in the spiral shock.  Our isothermal assumption further ignores complex heating and cooling within the ISM.  In reality, clouds will show variation in temperature between, e.g., the spiral arm and interarm regions.  
New simulations that implement heating and cooling prescriptions find dispersions in the range 3-7 km/s (\citealt{hopkins}; \citealt{dobbs2011}).  According to eq. (\ref{eq:prediction}), an error of this magnitude ($\sim$50\%) would correspond to an uncertainty of ~0.1 dex in the gas depletion time (25\% uncertainty in $\tau_{dep}$).  As this is the typical measurement uncertainty in $\tau_{dep}$ \citep{leroy08} our formalism can conversely accommodate ~50\% uncertainty in the velocity dispersion.  Given the current data quality, we feel that the isothermal assumption is valid for our purpose, although future observations (i.e. with ALMA) will make it possible to test such a scenario.  

Numerical calculations are clearly necessary to perform a detailed energy balance and follow the evolution and equilibrium of individual clouds during passage through the spiral arm.  
Still, the fact that we observe a correlation at all provides some indication that our simple expression is good to first order, even neglecting shocks and energy losses. 

The form of the correlation we consider here (plotting $\ln{\tau_{dep}}$ vs. $v_S^2/4\sigma^2$) by design represents the impact of dynamical pressure on the lengthening of the gas depletion time with respect to that of virialized clouds.  
But this ignores that, upon entering the arm, some clouds may themselves not be in virial equilibrium, i.e. due to non-zero streaming in the interarm.  
Even accounting for non-zero interarm streaming introduces little change to the correlation from environment to environment.  
We recalculated all fits including the inter-arm streaming velocities, i.e. according to eq. \ref{eq:full}, where now the lengthening in gas depletion time in the arm is relative to the depletion time in the interarm.  
The $\gamma$ and $\tau_{dep,0}$~associated with the best-fit slopes and intercepts in each of the five environments under consideration are listed in the middle part of Table \ref{tab:fits}. 
We find on average $\tau_{dep,0}$=0.66$\pm$0.26 and $\gamma$=-1.56$\pm$0.74 and a trend with environment that similarly resembles that obtained from direct fits to the mass spectra, although with slight exaggeration (see the blue points in the top and bottom panels in Figure \ref{fig:slopes_ints}).  
Given that the interarm velocities are smaller than the arm velocities, this adjustment is responsible for less deviation from the simplest case than when we impose a global power law mass spectrum.  
Estimates for $\tau_{dep,0}$ accounting for the inter-arm streaming are more similar to the values estimated including only arm streaming than to those where the slope is fixed to $\gamma$=-1.72 (see the entries in the bottom part of Table \ref{tab:fits} and plotted in Figure \ref{fig:slopes_ints}).  

\section{Discussion}\label{sec:discussion}

The close link between the predicted stable cloud mass and $\tau_{dep}$ described here strongly implicates the influence of dynamical pressure on the formation and evolution of the GMCs observed in M51.  
Indirect evidence for the role of surface pressure in cloud equilibrium is also seen in the lack of a clear size-linewidth relation for clouds in the arm (and interarm; \citealt{colombo2013}; Hughes et al. 2012b), which, when observed, is taken as evidence for virialized clouds.  
Our findings reiterate the warnings of \citet{shetty11} against the use of the virial parameter for determining cloud properties and quantifying boundedness.  
The external pressure--and the decrease in surface pressure in the presence of strong streaming motions in particular--must be taken into account.   

The cloud-decomposed emission in the PAWS data cube shows other consistencies with the predictions for the impact of dynamical pressure.  \citet{colombo2013} present evidence that the GMCs in the arms are more massive than those in the interarm by about an order of magnitude, clear by the offset in the mass spectra for clouds in the two zones.  
Meanwhile,  a larger fraction of the inter-arm gas is in the form of clouds than in the spiral arm \citep{colombo2013}.  
In our picture, regions with high streaming velocities should have fewer collapsing clouds, and the few that exist will be of higher mass than clouds that are virialized.  

At present, the tight link between gas flows and star formation is based on only one galaxy, and tests of similar systems will be necessary to firmly establish the role of dynamical pressure.  
In this discussion section, we compare dynamical pressure with other sources of cloud stability and highlight several aspects of star formation that can be explained uniquely by our picture. \\

\subsection{Other potential sources of cloud stability}\label{sec:stability}

In the previous section we revealed an environment-dependent correlation between gas depletion time and dynamical pressure.  
This arises from the dependence of the star formation efficiency on the mass fraction of collapse unstable clouds per free fall time, which varies with location in the disk; the mass fraction of collapse unstable clouds is specified by the environment-varying cloud mass spectrum index, together with the stable mass threshold set by the dynamical pressure.  
In our picture, dynamical pressure effectively stabilizes the clouds and thus prevents all but those above the raised stable mass threshold from collapsing to form stars.  
But there are several other potential sources of cloud stability.  We review these here and consider the possible role shocks might play in stimulating the observed pattern of star formation.  
We conclude that none of these cases can adequately explain the observed pattern in the SFE or gas depletion time, leaving dynamical pressure as the most compelling source of stability.  

\begin{figure}[t]
\begin{centering}
\includegraphics[width=1.1\linewidth]{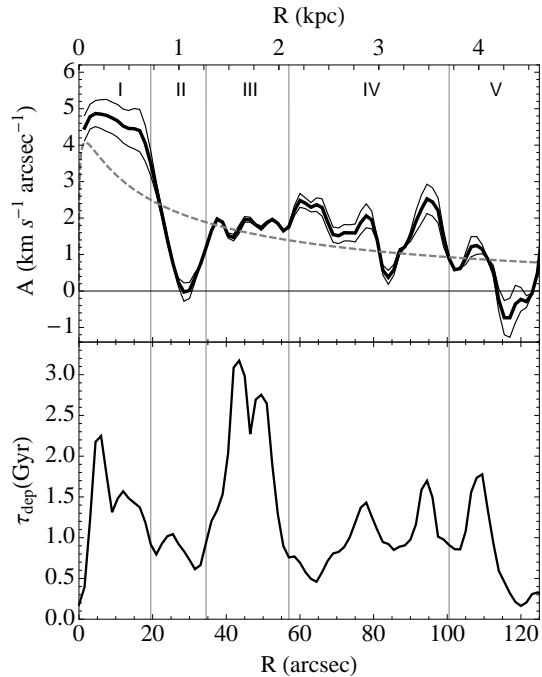}
\caption{(Top) Profiles of the background shear due to differential rotation (gray dashed line) and the shear in the bar and spiral arm regions, calculated from our measure of $v_\phi$ in Figure \ref{fig:stream} (black curve).  Thin black lines represent the rms dispersion in solutions with $\pm$5$\degree$ variation in the assumed pitch angle $i_p$. 
(Bottom) For reference, the radial trend in the molecular gas depletion time is depicted in solid black, repeated from Figure \ref{fig:torqsfe}.  \label{fig:shearplot}
}
\end{centering}
\end{figure}
\subsubsection{Shear}
\label{sec:stabShear}
As the locations of strong non-circular motions and enhanced surface density, spiral arms are preferentially sites of reduced shear (as argued by \citealt{elmegreen87}; \citealt{elmegreen94} and see \citealt{kimOstriker02}; \citealt{kimOstriker06}).  
This also makes them favored sites of star formation since, in the absence of shear to counter self-gravity, gas can form coherent cloud structures that become gravitationally unstable.  

The link between enhanced shear and reduced star formation has been studied by \citet{Seigar} (and see \citealt{Weidner} and \citealt{HocukSpaans}).  
While this may hold on global scales, Figure \ref{fig:shearplot} suggests otherwise for M51.  The link between reduced shear and star formation does not hold uniformly along the spiral arms on two counts: 
1) Shear as measured by the Oort parameter A =1/2($v_{\phi}$/R - $d$$v_{\phi}$/$d$$R$) in the spiral arm is very clearly non-zero along the length of the arm.  
Shear in the spiral arm only counters the local background shear (from the rising rotation curve; see $\Omega$ in Figure \ref{fig:speeds}) between 18\arcsec$\lesssim$$R$$\lesssim$28\arcsec and further out near $R$$\sim$85\arcsec and $R$$\sim$100\arcsec.  
In these outer regions, in particular, it appears that shear is responsible for the lengthening of the depletion time.    2) The region 35\arcsec$<$$R$$<$60\arcsec ~where we find a pronounced increase in $\tau_{dep}$ (or decrease in SFE) is characterized by a similar, if not lower, degree of shear as in the neighboring zone at larger galactocentric radius, which is characterized by relatively more star formation.  Moreover, as examined in Appendix \ref{sec:appxCritdens}, the threshold for cloud instability and collapse in the presence of shear (as well as Coriolis and tidal forces) is exceeded by the observed gas surface density everywhere along the length of the arms.  
To prevent star formation on cloud scales between 35\arcsec$<$$R$$<$60\arcsec ~another source of stability is clearly required.  

Our finding echos that of \citet{dib2012} who considered the role of shear on the scale of individual molecular clouds in the Galactic Ring Survey.  
In all cases, they find that shear does not balance gas self-gravity and conclude that turbulent motions, stellar feedback\footnote{Star formation driven winds from successive star formation events are thought to deposit the energy required to sustain turbulence (which otherwise dissipates quickly; i.e. \citealt{murrayQT})} and/or magnetic fields are the principle agents of cloud stability.  

\subsubsection{Turbulence and Magnetic Fields}
To produce the observed pattern in $\tau_{dep}$ in M51 we might similarly expect turbulent motions to be enhanced within clouds in the zone 35\arcsec$<$$R$$<$60\arcsec ~compared to clouds from neighboring zones.   
Our high resolution PAWS observations, where the line-of-sight velocity profiles sample turbulent velocities above $\sim$2 km s$^{-1}$ (\citealt{petyPAWS}; \citealt{colombo2013}) on cloud scales.  
do not show such a pattern, however.  
The velocity dispersion shows no pronounced change along the arms (see Figure \ref{fig:stream}).  The virial parameter $\alpha$, which measures the balance between the internal kinetic energy and gravity of clouds,  shows no significant difference between this region of the spiral arm and the neighboring segment, beyond $R$=60\arcsec ~\citep{colombo2013}.  
Either turbulent motions are not responsible for the radial variation in $\tau_{dep}$ shown in Figure \ref{fig:torqsfe} or they are present but fall below our velocity resolution.  Below we will consider shocks and their influence on cloud velocity dispersions in more detail.  
Here we note that, given the marked absence of star formation in this zone, cloud support can not come from turbulence driven by stellar winds/feedback.   

Magnetic fields are another potential influence on the organization of the ISM, but their role in cloud stabilization is not clear (or observationally established) at this point.  Since they should pervade the ISM and clouds equally, it seems unlikely that they are responsible at the cloud level for the trends we see here.   

\subsubsection{Shocks}
Spiral arm shocks, on the other hand, have been long appreciated for their role in the formation of GMCs and the  triggering of star formation (e.g. \citealt{rs87}; \citealt{gittins03}; \citealt{gittinsClarke}; \cite{kimOstriker02}; \citealt{dobbs08}).  Shocks bring the gas surface densities high enough to initiate gravitational instability and at the same time favor massive GMC formation via collision and agglomeration.  
Clouds formed as a result of shocks are predicted to have elevated velocity dispersions \citep{bonnell}, with clear implications for cloud boundedness and hence the global patterns of star formation.  

In this shock picture, variations in cloud properties would also presumably emerge {\it along} the spiral arms depending on the properties of the shock, or as a result of variations in the balance between shear, self-gravity and agglomeration (as well as the properties of the pre-shock medium, i.e. \citealt{dobbsbonnell}; \citealt{dobbs08}).  
For example, the zone we identified with very little star formation might preferentially host unbound clouds formed via agglomeration in the shock rather than gravitational instability.  
However, there are several observations that counter this interpretation, as summarized below.  

First, the surface density in this region is among the highest along the arms.  In comparison to the other zones with star formation--and which presumably host bound clouds--this zone should likewise favor cloud formation via gravitational instability.   

At these high surface densities, unbound clouds might still emerge if agglomeration builds on the masses of clouds already formed by the dominant process of instability in self-gravitating gas (as realized in the simulations of \citealt{dobbs08}).  
The properties of these clouds are predicted to depend on the strength of the shock, with stronger shocks leading to the accumulation of more massive, unbound clouds (\citealt{dobbsbonnell}; \citealt{dobbs08}).  

\begin{figure}[t]
\begin{centering}
\includegraphics[width=1\linewidth]{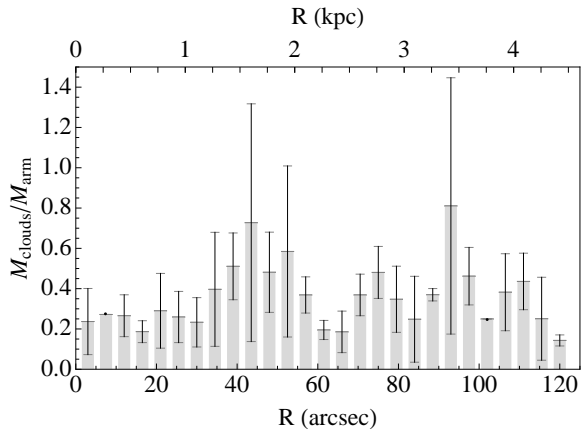}\\
\end{centering}
\caption{
Histogram of the ratio of the molecular mass in catalogued clouds $M_{clouds}$ (estimated from the cloud CO luminosity; see \citealt{colombo2013}) to the total (luminous) mass in the arm $M_{arm}$ measured in 4.5\arcsec ~radial bins. Thin lines delineate the dispersion in cloud masses within each bin.  
}
\label{fig:massratio}
\end{figure}

This case could potentially mimic a varying gas depletion time since, from this point, stars formed from individual bound clouds embedded in a larger unbound structure do so at an overall low efficiency.  
We might therefore expect the inverse of $\tau_{dep}$ (the SFE) to be well-matched to the ratio of total mass in clouds to the total mass in the arm (i.e. a low mass of potentially star-forming material where $\tau_{dep}$ is long).   
But Figure \ref{fig:massratio} shows no such behavior.  Instead, the mass in clouds relative to the total mass seems even elevated where the star formation is preferentially low.  
This is very strong evidence that the agglomeration of unbound cloud associations via shocks is not responsible for the observed pattern in $\tau_{dep}$, especially since it conservatively assumes that all catalogued clouds will collapse to form stars. 

As analyzed by \citet{colombo2013}, the shape of the cloud mass spectrum provides a record of the processes responsible for the formation (and destruction) of clouds in M51.  
GMCs in the spiral arms are predominantly formed by gravitational instability, although agglomeration plays a role in organizing the ISM into cloud associations along these structures.  
Notably, the cloud mass spectrum shows no significant change in overall shape along the arm \citep{colombo2013}.  This suggests that the mix of instability and agglomeration in cloud formation is equal in arm environments, namely between the spiral segment neighbored by star formation on its convex side and the segment 30\arcsec$<$$R$$<$60\arcsec with little to no star formation.  
The pattern in star formation therefore does not appear to be tied with cloud formation mechanism.  
\subsubsection{Triggered star formation}
Star formation triggering by spiral-arm shock dissipation is also very clearly not responsible for the observed pattern in the SFE.  
Since stronger shocks are accompanied by larger velocity gradients and thus dissipate more energy, we might have expected relatively short gas depletion times in exactly the locations of strongest spiral shock (i.e. anywhere away from corotation), unlike what is observed.  This picture would also require the shock strength to vary from one segment of the arm to the other (i.e. 40-60\arcsec ~compared with 60-80\arcsec) to explain the difference in the star formation properties between these two zones, but this does not appear to be the case \citep{schinner2013}.  Taking the spiral shock strength as the size of the offset between the molecular spiral arm from the potential minimum traced by the peak in the old stellar light distribution imaged at 3.6 $\mu m$, \citet{schinner2013} find only a modest difference between the zones 40-60\arcsec ~and 60-80\arcsec.  
We therefore conclude that, although spiral-arm shocks may be active in M51, it is not clear that they have a strong, direct influence on the triggering of star formation.  

\begin{figure*}[t]
\begin{centering}
\begin{tabular}{cc}
\includegraphics[width=.51\linewidth]{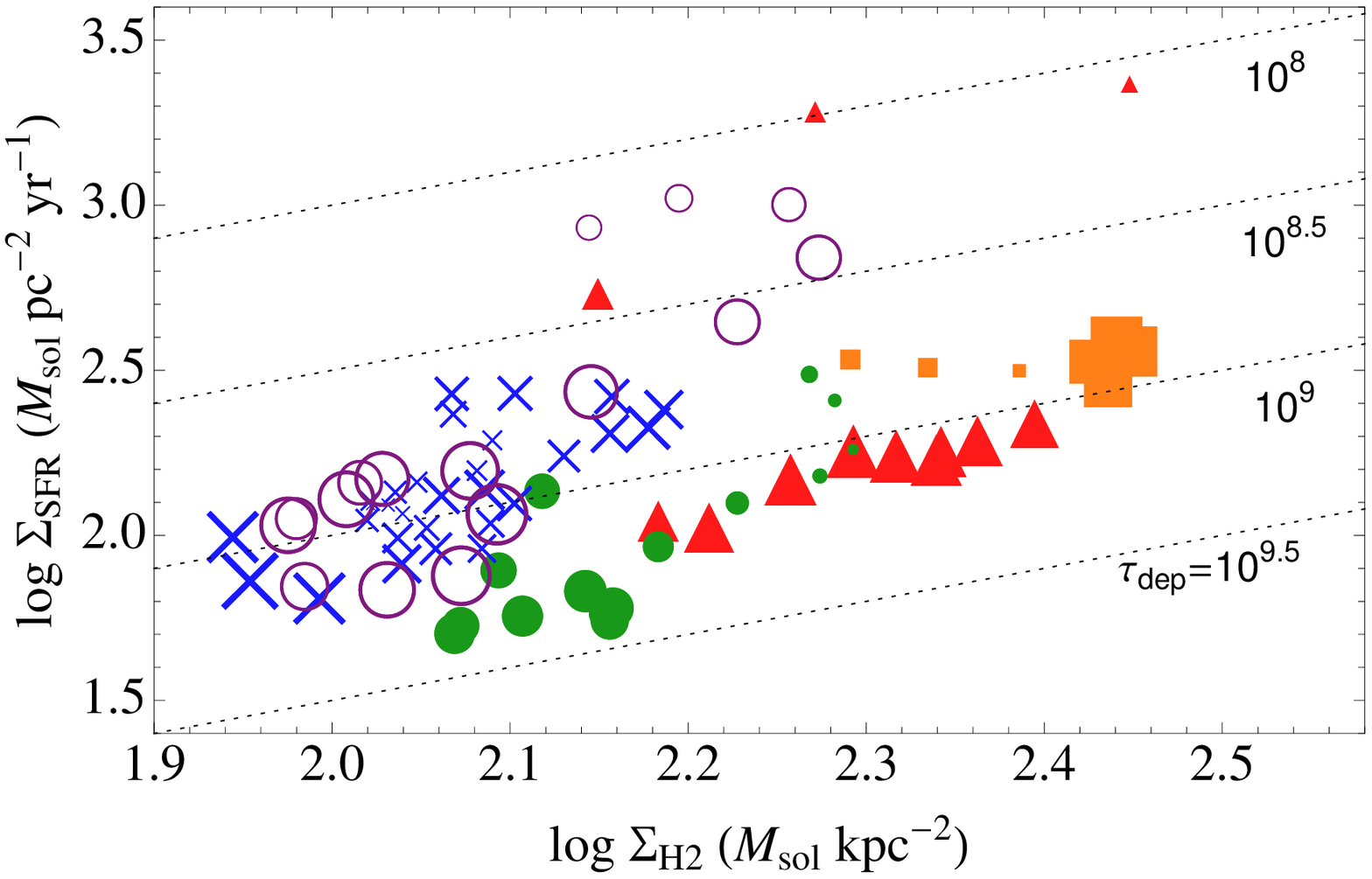}&
\includegraphics[width=.51\linewidth]{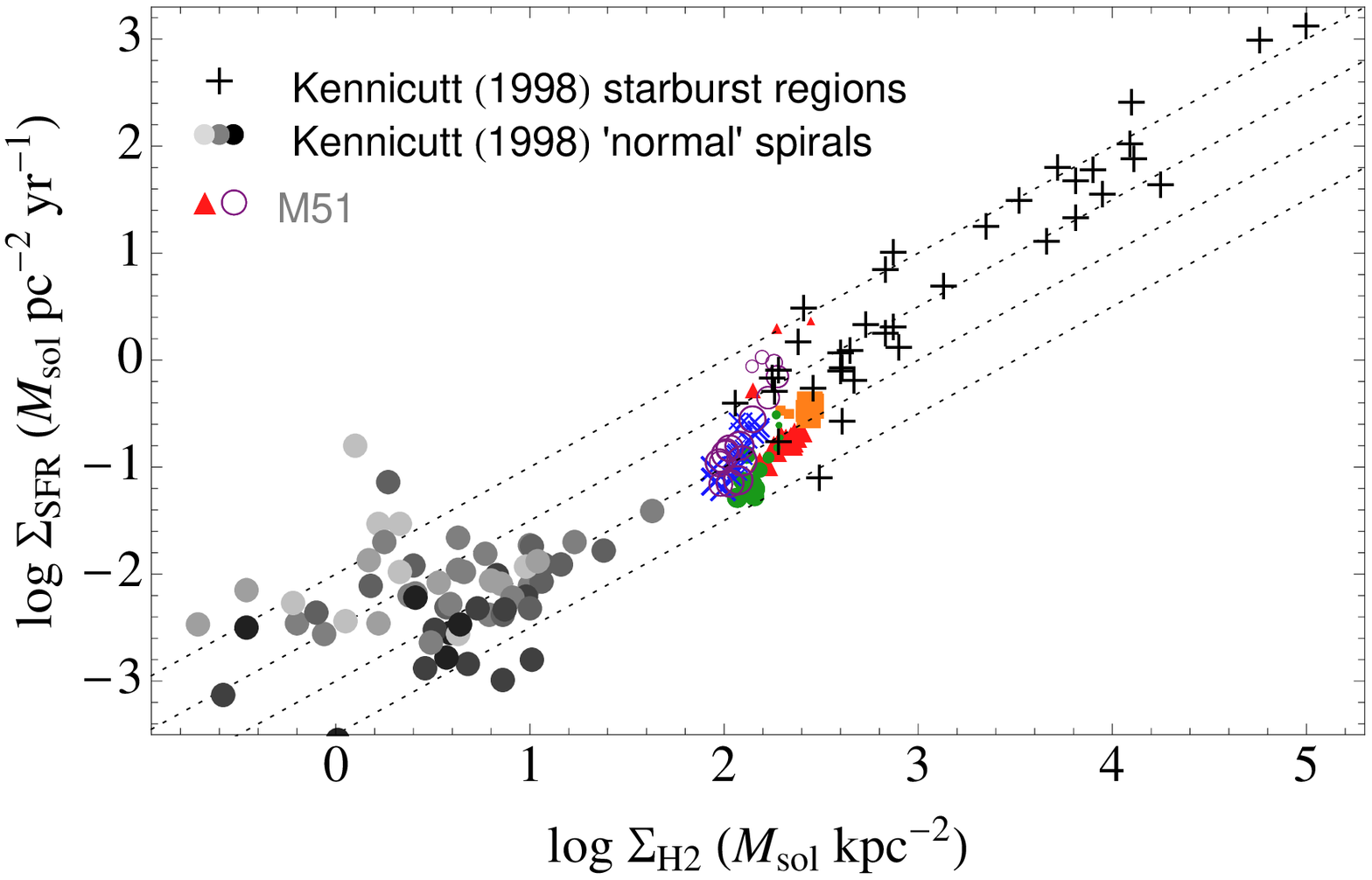}\\
\end{tabular}
\caption{
(Left) Star formation surface density $\Sigma_{SFR}$ versus observed molecular gas surface density $\Sigma_{H2}$ measured in 2.4\arcsec ~radial bins from the center to the edge of the PAWS field, with point colors and shapes as in Figure \ref{fig:velsfe}.   The size of each point is scaled to the kinematic term $v_S^2/(4\sigma^2)$ defined in $\S$\ref{sec:dynamicalPquant}.  Dashed lines correspond to constant gas depletion times 10$^{8.5}$, 10$^9$, 10$^{9.5}$ and 10$^{10}$yr.   
(Right)  Star formation surface density $\Sigma_{SFR}$ versus observed molecular gas surface density $\Sigma_{H2}$ where points represent local measurements in M51 (colored points, repeated from the plot at left) and data from \citet{kennicutt98} for a sample of starbursts (black crosses) and `normal' spirals (filled circles; grayscale indicates disk morphology, from (black) early-type to (light gray) late-type).  Lines of constant depletion time are repeated from the plot on the left.   
\label{fig:ksrelation}
}
\end{centering}
\end{figure*}

\subsection{Implications for patterns of star formation in observed and simulated galactic disks\label{sec:disc_patterns}}

Up to this point, simulations and semi-analytical models of GMC formation and evolution rarely consider non-axisymmetric potentials, and so streaming motions and dynamical pressure have been neglected.  
Variations in the star formation efficiency and cloud stability have mostly focussed on the role of turbulent support and turbulent energy dissipation (\citealt{krumholzmckee}; \citealt{murrayQT}; \citealt{feldmann}).  The few simulations with fixed spiral potentials have thus far focused only on very tightly-wound spirals with pitch angles $i_p$$<$10$\degree$ and so have not probed the regime of strong streaming motions.  
In addition, these simulations consider the Bernouli principle more for its role in spur formation via the Kelvin-Helmholz instability, which has subsequently been found ineffective in real galaxies compared to the magneto-Jeans instability \citep{kimOstriker02}.  

From the observational perspective, we can already infer the role of dynamical pressure in galaxy disks.  
The strong dynamical dependence of the conversion of gas in to stars presumably contributes to the frequent lack of a clear signature in the angular cross correlation between gas and young star tracers (\citealt{foyle11}; \citealt{egusa}). 
The angular offset between the young stars and gas predicted as a result of the propagation of the spiral pattern (combined with the timescale for star formation; e.g. Roberts \& Stewart 1987), will not be observed, even where the gas column is high, if the gas is dynamically prevented from forming stars.

The influence of dynamical pressure is also manifest at the global level.  We suspect that it keeps stronger grand-design spirals from forming an excess of stars relative to weaker, flocculent spirals with similar global molecular gas surface densities, as observed (see Elmegreen \& Elmegreen 1986).   
Whereas stronger spirals result in locally higher gas surface densities, and might therefore be expected to exhibit higher star formation rate surface densities, they also locally stabilize clouds through dynamical pressure.  
These two effects presumably together keep the global star formation rate independent of spiral strength.   
An increased velocity dispersion in the strongest spiral arms will achieve the same result, but we note that  a distinction from dynamical pressure is difficult to establish, as observations at resolutions above the size scale of GMCs will also reflect unresolved streaming motions.

The strong internal gas flows associated with galaxy interactions suggests that dynamical pressure may be a common feature in such scenarios.  M51 and its companion are an ideal test-bed for exploring the impact of interactions on gas organization and subsequent star formation.  Our consideration of the impact of M51s companion galaxy has been so far indirect, concerning the non-axisymmetric structure across the PAWS field that the interaction helped to stimulate and shape. Features that are more directly influenced by the companion, namely the ``bridge" that extends between the two galaxies and the opposite ``tail", are beyond the edge of the PAWS field of view, precluding the study of the ISM organization and the properties and progression of star formation with the same detail as we have undertaken at smaller galactocentric radius. As the ISM composition is changed at large radii, as well, future insight in to the nature of the impact of the interaction will require extended, high resolution kinematic tracers of the molecular and atomic ISM phases. 

\subsubsection{Relevance for barred galaxies}
The scenario we find here -- low star formation efficiency in the CO-bright inner spiral arm segments -- can be thought of as comparable to the dust lanes along the length of bars typically observed to harbor little star formation (\citealt{sheth05}; \citealt{comeron}).  
Star formation is thought to be inhibited in the strongest bars because of strong shocks and high shear (\citealt{athanassoula92}; \citealt{sheth05}; but see \citet{zurita} who suggest shocks help trigger star formation).   
In our picture, on the other hand, the strong radially inward streaming motions that accompany strong bar torquing  stabilize clouds, shutting off star formation.  
We find this to be a compelling interpretation since the two conventional kinematic sources of cloud stabilization -- enhanced shear and shock-induced turbulent motions -- can not be active in the comparable zone in M51 with high $\tau_{dep}$ ($\S$ \ref{sec:stability}).   
Not only is shear reduced in this zone, but the gas is not characterized by extraordinary line-widths, implying that the fewer collapsing clouds in this zone are not the result of preferentially enhanced turbulence.   

Dynamical pressure can also be invoked, as an alternative to shear and shocks, to explain the suppression of star formation in the Galactic center \citep{longmore}.  The observed gas kinematics in the Milky Way's central molecular zone (CMZ) are consistent with 
characteristic motions along bar orbits, but the evidence for inflow motions along a bar shock are less clear; the fall-off in the signature `parallelogram envelope' with increasing distance from the center in the pv diagram \citep{binneycmz} 
 is not especially obvious.  This could be due to the data quality (sensitivity issues), or it may indicate that the weak oval nuclear bar does not sustain shear and shocks.\footnote{Note that shear is expected to be zero in the absence of shocking due to gravitational or viscous torques on the gas within the zone of the bar.}  In this case, if shear is not present to stabilize clouds, dynamical pressure due to non-circular motions in the bar may be responsible for altering the ability of clouds to form stars. Estimates of the critical surface densities for shear and dynamical pressure (see Appendix \ref{sec:appxCritdens}) could help further establish the likelihood of these two scenarios.  

\subsection{Implications for local and global star formation relations\label{sec:disc_SFrelation}}

As demonstrated in Section \ref{sec:emp_corr}, the influence of dynamical pressure manifests itself by introducing departures from a fiducial gas depletion time $\tau_{dep,0}$, with higher streaming motions leading to longer $\tau_{dep}$.  
According to the formalism there, this fiducial $\tau_{dep,0}$ is representative of virialized clouds, and our fitting in the case of M51 ($\S$ \ref{sec:emp_corr}) suggests that $\tau_{dep,0}$ can be as low as $\sim$0.2Gyr.  
By analogy with \citet{krumholz12}, we view $\tau_{dep,0}$ as set by the cloud free fall time $t_{ff}$ modulo some roughly constant dimensionless efficiency $\epsilon_{ff}$ ($\sim$0.01 in a supersonically turbulent medium; \citealt{krumholzmckee}) measuring the intrinsic efficiency of star formation in a cloud.  But this efficiency is modulated by the dynamical pressure term so that 
\begin{equation}
\tau_{dep}=\frac{t_{ff}}{\epsilon_{ff}}e^{-(\gamma+1)v_{s}^2/4\sigma^2}\label{eq:depl}
\end{equation}   
This change to the efficiency reflects the influence of the environment of a cloud on its star formation properties.   
Alternatively, we can view the exponential term as altering the time required to complete collapse. 
This would be more similar to the \citet{krumholz12} picture, in which the free-fall time depends on the cloud 'regime'; clouds that are dynamically coupled to their environment with a low density contrast have shorter depletion times than clouds that are decoupled, at high overdensity. 
The environmental dependence parameterized by this bimodal free-fall time leads to a successful fit to observations ranging from star forming clouds in the Milky Way to high-z starbursts.  
Below we show that our predictions can also explain the observations by presenting a form of the star formation law that leads to analogous environment-dependence.  
In contrast to the \citet{krumholz12} picture, our environmental dependence is smoothly parameterized by the dynamical pressure, which affects the coupling to the environment by reducing the cloud surface pressure.  

\subsubsection{Impact on local trends}
With the change in the depletion time in eq. (\ref{eq:depl}), the star formation relation can be expressed as
\begin{equation}
\Sigma_{SFR}=\frac{\Sigma_{gas}}{\tau_{dep,0}}e^{(\gamma+1)v_{s}^2/4\sigma^2}
\end{equation}
We see immediately that 
the range in streaming velocities both within and among spiral galaxies should contribute to the scatter in the Kennicutt-Schmidt star formation relation (i.e. \citealt{leroy}; \citealt{bigiel}).\footnote{Pixel-by-pixel comparisons between $\Sigma_{gas}$ and $\Sigma_{SFR}$ should reveal such departures from a universal gas depletion time most clearly;  the $\tau_{dep}$ measured from larger-area averages is the true $\tau_{dep}$ weighted by $\Sigma_{SFR}$ and will thus less strongly track low star formation environments.}    
This is illustrated in Figure \ref{fig:ksrelation}.  The left panel shows measurements from annuli within M51, while these annuli are plotted together with integrated measurements from \citet{kennicutt98} for normal spiral galaxies and starburst environments on the right.  

Regions in M51 with the highest streaming motions sit at the longer gas depletion times, falling below the points with relatively less streaming.  Spiral galaxies with similar non-circular motions will show a similar spread in (local) depletion time.  
But more importantly, given that typical maximum streaming velocities are 10-15 km s$^{-1}$, we expect that spiral galaxies will preferentially show roughly the same maximum $\tau_{dep}$, especially since measurements on the scales of 1 kpc and larger tend to be weighted to the bright, high density spiral arms.    
Galaxies with little to no streaming, on the other hand, will show lower depletion times; the gas in these systems will form stars closer to the fiducial gas depletion.  

In this light, we might view the gas depletion time $\tau_{dep}\sim$2 Gyr measured by \citet{bigiel} for a sample of spiral galaxies (ranging in stellar mass from $\sim$10$^7$ to 10$^{11}$ $M_\odot$) as characteristic of the depletion time in the presence of spiral streaming motions.  
At the same time, we can attribute the lower $\sim$1~Gyr gas depletion times in the \citet{kennicutt98} starbursts and also characteristic of low-mass systems assuming a Galactic conversion factor $X_{CO}$ (\citealt{leroy2013}; but see discussion about the impact of $X_{CO}$ below) to the fact that this star formation is occurring in a relatively stable dynamical environment.  
In the former case, the star formation typically occurs at resonances where the streaming motions are zero, while in the latter case, streaming is absent because these systems fail to develop the non-axisymmetric structure that drives these motions (the low mass means the disk never becomes unstable).  
\subsubsection{Impact on global trends }\label{sec:globalSF}
In reality, we expect to see a range of gas depletion times intermediate between these cases.  
Since disks of a certain mass will be unstable to structure of a particular wavenumber and pitch angle $i_p$, as upheld in the density wave paradigm, we can express the $i_p$ dependence in terms of galaxy mass, or circular velocity $V_c$, using the dispersion relation for $m$-armed density wave spirals (i.e \citealt{bt}).  We use the perturbed continuity equation
\begin{equation}
v_r\approx \frac{(\Omega-\Omega_p)\tan{i_p}}{m}\frac{\Sigma}{\Sigma_0}  \label{eq:cont}
\end{equation}
to express the angular velocity difference $\Omega-\Omega_p$ in terms of the radial streaming velocities $v_r$, where $\Omega_p$ is the wave/spiral pattern speed and $\Sigma$ ($\Sigma_0$) is the perturbed (unperturbed) gas surface density.
In this way, we find
\begin{equation}
\tan{i_p}=\frac{1}{4\pi V_c^2}\left[v_r^2\left(\frac{\Sigma}{\Sigma_0}\right)^2-\sigma^2\right]
\end{equation}
where $V_c$=$\Omega R$ is the rotational velocity.  
This reproduces the well-known observation that spiral arm pitch angle increases with galaxy mass, $V_c$, or Hubble type \citep{kennicuttPitch}.  

Now from eq. (\ref{eq:cont}) we predict that the magnitude of the streaming motions
\begin{equation}
v_{S}\approx v_{r}\propto \frac{V_c}{m}\frac{\Sigma}{\Sigma_0}  \label{eq:vsimp}
\end{equation}
to lowest order in $V_c$.  (Note that this applies globally, away from corotation, so that $v_r$$\sim$$v_{\phi}$ and $v_{S}$$\sim$$v_r$.)  

Less-tightly wound spirals in the more massive, early type disks drive stronger streaming motions. This implies that spiral galaxies with higher mass (larger $V_c$) should exhibit longer gas depletion times.   
This is in fact what global measurements of $\Sigma_{SFR}$ and $\Sigma_{gas}$ suggest: the normal spiral sample in the KS plot on the right in Figure \ref{fig:ksrelation} indicates an ordering between depletion time and morphology/Hubble type.  
(Note that while the starbursts lie above M51 and the earlier-type spiral galaxies, they are more similar to the lower mass later-type galaxies, which exhibit similarly short depletion times.)  
This is even clearer in the HERACLES sample of galaxies, where integrated measures of the gas depletion time  smoothly decline from early to late spiral types (\citealt{leroy2013}).  

This dependence of $v_S$ on $V_c$ accounts for the observed weak correlation between $\tau_{dep}$ and stellar mass among galaxies observed by \citeauthor{leroy08} (\citeyear{leroy08}, \citeyear{leroy}) and \citet{saintonge}.  
According to eq. (\ref{eq:cont}) or (\ref{eq:vsimp}), the scatter in that relation at fixed stellar mass can be attributed to, e.g., morphologic variations (number and strength of arms, pitch angle) and gas content (see below).  
It can also explain the modest success of the `dynamical Schmidt' law in fitting observations, like other similar prescriptions that express the gas depletion time in terms of $V_c$ (i.e. \citealt{tan2010} and references therein).  
Here, though, we predict a non-linear dependence $\tau_{dep}\propto e^{-V_c^2/\sigma^2}$ that can be approximated by $V_c$ only at low $V_c$ (and in particular at low $V_c$/$\sigma$).   
\subsubsection{High z star formation relation}
Our form of the star formation relation is also capable of linking local and high-z star formation. Observations suggest that star forming galaxies at earlier times form stars at high efficiency compared to their local counterparts (i.e. \citealt{daddi}), with depletion times comparable to those in local starbursts environments.  
(Local and $z$=2 star forming disks fall on a non-linear KS relation with power-law index $n$$\sim$1.3, e.g. \citealt{daddi}; \citealt{krumholz12}.)  
High-z starbursts and mergers are inferred to consume their gas even faster, giving the appearance of a bimodality in the KS relation at $z$$\sim$2 (\citealt{daddi}; \citealt{genzel}).  These authors suggest that these two modes of star formation can be linked by adopting a form for the star formation relation that incorporates the dynamical time, i.e. letting the the dynamical time set the gas depletion time.  However, \citet{krumholz12} demonstrate that this form of the star formation relation can not be consistently applied to local systems.  
Instead, the fiducial gas depletion time is argued to be lower in high-z starbursts likely due to the expected high densities that result under the compressive weight of the ISM.  
As is clearer below, our picture presents a parameterization of this weight in terms of $V_c$/$\sigma$ or the gas mass fraction $f_g$.  

To more clearly illustrate the differences we expect to see between systems at low and high $z$, we alternatively express the streaming velocity in terms of the Toomre $Q$ and gas-to-total mass ratio $f_g$.  
Using that 
\begin{equation}
Q=\frac{\sqrt{2(\beta+1)}\sigma\Omega}{\pi G \Sigma_{gas}},
\end{equation}
where $\beta$=$d$ln$\Omega$/$d$ln$R$=1 for solid body rotation and $\beta$=0 in the case of flat rotation curves, 
$V_c^2$$\approx$$\pi G \Sigma_{tot}R_{edge}$, and assuming that the stellar and gaseous disks cover roughly the same area so that $f_g$=$\Sigma_{gas}/\Sigma_{tot}$ then, together with eq. (\ref{eq:vsimp}), we find  
\begin{equation}
\frac{v_{S}}{\sigma}\propto \frac{\Sigma}{\Sigma_0}\frac{1}{m}\frac{\sqrt{2(\beta+1)}}{f_g Q} \label{eq:highz}
\end{equation}

Star forming disks at high-z have lower gas depletion times than similarly marginally unstable disks at $z$=0 due to the higher gas fraction at earlier times.  
High-z starbursts and merging systems can have even lower depletion times, given the likely higher gas fractions.   

Note that the observed trends also follow from eq. (\ref{eq:vsimp}), as the ratio $V_c$/$\sigma$ is lower in star forming disks at earlier times, and even lower in starbursting mergers (i.e. LIRGS/ULIRGS), given the enhanced turbulence and larger $\sigma$.  
This may explain why a linear dependence of $\Sigma_{SFR}$ on $V_c$ can be used to fit observations at high z, where the quantity $V_c$/$\sigma$ is intrinsically low, and thus a suitable approximation of the exponential dependence we find (see $\S$ \ref{sec:globalSF}).    
We emphasize, though, that $V_c$ is only part of the description, as the ratio $V_c$/$\sigma$ provides the more fundamental measure of the importance of dynamical pressure.    
As described in $\S$ \ref{sec:dynamicalPqual}, even when streaming motions are present, dynamical pressure is not important if clouds can equalize before they undergo translation by more than a cloud length, that is, if $\sigma$$>$$v_S$.  
With eq. (\ref{eq:vsimp}), this is equivalent to the condition $\Sigma_{gas}$$\sim$$\Sigma_{tot}$ for gas disks in hydrostatic equilibrium (recall that by writing eq. (\ref{eq:vsimp}), we replace the existence of streaming motions by the requirement for a disk massive enough for instabilities to form).  
So even in the presence of streaming motions, a high gas fraction prevents a lengthening in the depletion time.   

Consider, for example, the difference between starbursts at low and high z, which each have short depletion times but very different $V_c$.  
In the case of local starbursts, the high gas densities (and high star formation) are possible because of the absence of motion relative to the background rotating potential; locally motionless gas (fed from neighboring regions) can retain and build a molecular reservoir.  
This scenario is different from the conditions that lead to bursts of star formation at higher redshift, where accreted external gas builds up globally high surface densities and streaming motions themselves can be quite high (i.e. given the high measured $V_c$; \citealt{genzel}).  
Even though these streaming motions are present, dynamical pressure is less important, according to the measured ratio $V_c$/$\sigma$.  

\subsubsection{Comparison with other findings}
In our picture, shorter depletion times are associated with greater levels of turbulence and higher $\sigma$.  This is also a feature of the model proposed by \cite{renaud} based on feedback-regulated turbulence. 
The higher turbulence in $z$=2 mergers as compared to $z$=2 disks leads to a higher fraction of (locally) dense gas, higher overall star formation rate and thus a vertical offset in $\Sigma_{SFR}$ vs. $\Sigma_{gas}$ space.
Despite consistency with this dependence of $\tau_{dep}$ on $\sigma$, though, we interpret the role of $\sigma$ differently.  In the \citet{renaud} model, lower levels of turbulence lead to less effective turbulent-triggering of star formation and longer depletion times.  In our picture, a lower $\sigma$ leads to a more destabilized system and thus to the onset of streaming motions, which lengthen the depletion time.  Note that both of these pictures appear to imply that star formation consumes gas quicker when the gas is more stabilized (according to the increase in Toomre Q with the higher effective gas sound speed in the presence of greater turbulence), perhaps contrary to expectations.  

Interestingly, at fixed $\sigma$, eqs. (\ref{eq:depl}) and (\ref{eq:highz}) together predict that the gas depletion time increases nonlinearly with decreasing gas fraction, or that the star formation efficiency increases nonlinearly with gas fraction.  
In addition, more centrally concentrated gas, situated in the portion of the disk where the rotation curve is rising and $\beta$$\sim$1, will have slightly longer $\tau_{dep}$ than more extended disks with $\beta$=0.  
In the former case, shear prevents gas from forming collapsing structures.  Our picture is thus also able to describe the findings of \cite{saintonge} and may prove descriptive, from an observational perspective, of `morphological quenching' as envisioned by \citet{Martig}.

We can now put this all together to predict a dependence  
\begin{equation}
\log\Sigma_{SFR}=\log\Sigma_{gas}-\log{t_{ff}/\epsilon_{ff}}-\frac{(\beta+1)}{4f_g}
\end{equation}
for a Toomre unstable disk (Q$\approx$1), and taking for simplicity $\Sigma/\Sigma_0$$\sim$$m$=2 and $\gamma$=-1.5 for the cloud mass spectrum index (i.e. \citealt{ros2005}). 
Assuming a universal $t_{ff}/\epsilon_{ff}$ that sets the intercept of a 'fiducial' linear relation between $\log \Sigma_{SFR}$ and $\Sigma_{gas}$, then   
systems will fall progressively below this line the smaller their gas fraction.  
This form of the star formation relation can consistently describe the strong offsets between various populations in the standard KS plot as assembled and presented by \cite{krumholz12}.  Systems with the highest gas fractions -- high-z starbursts, local starbursts, and even including star-forming clouds in the Milky Way -- are offset to the shortest depletion times.  Below these fall the high-z star forming disks at intermediate $\tau_{dep}$, followed by the local star forming galaxies at the longest $\tau_{dep}$.

Of course, streaming motions may not be responsible for the entirety of the offset in observed gas depletion time both within and among galaxies.  Some part of the offset can be attributed to, and removed by,   
variation in the CO-to-H$_2$ conversion factor ($X_{CO}$).  The bimodality between local and high-z systems is arguably enhanced due to the adoption of a bimodal $X_{CO}$ \citep{narayanan}.  
Allowing for a dependence of $X_{CO}$ on metallicity, low-mass dwarfs (with high $X_{CO}$) appear to have similar gas depletion times as more massive disks (\citealt{schruba12}; unlike in Figure \ref{fig:ksrelation} where a single $X_{CO}$ is assumed).  

On the other hand, it seems that not all depletion time variations between different populations can be minimized in this way.  
In M33, a spiral galaxy intermediate in mass and metallicity between the MW and Local Group dwarfs, the depletion time is shorter than for spirals in the \citet{kennicutt98} sample, even adopting a slightly higher than Galactic $X_{CO}$ appropriate for its metallicity \citep{braine2011}.  
In addition, although \citet{leroy2013} find that most variation in $\tau_{dep}$ internal to galaxies can be accounted for by varying $X_{CO}$, the differences in $X_{CO}$ between the disks and central regions of galaxies measured by \citet{sandstrom} in fact underscore the existence of two different `modes' of star formation \citep{leroy2013}.  
\section{Concluding Remarks}
 
In this paper, we examined the influence of non-axisymmetric stellar structure (i.e. nuclear bar, spiral arms) on gas flows, molecular cloud properties and star formation in the inner disk of the
iconic spiral galaxy M51.   Leveraging our unique view of gas motions, which includes both measurement of present-day torques and non-circular streaming motions decomposed from the line-of-sight velocity field, we establish an anti-correlation between gas flows and
strong star formation in M51.  To explain the observed gas flow and star formation patterns we developed a simple model, in which dynamical pressure
is a critical parameter for determining cloud stability against
gravitational collapse.  We report the following results and conclusions: \\

1. A radial profile of the stellar torques across the PAWS field
reveals distinct dynamical zones within which gas is driven radially inwards
or outwards in response to the non-axisymmetric structure present in M51's
inner disk.  
In the nuclear bar region ($\lesssim20\arcsec$) and inside
the corotation resonance of the first spiral arm ($35\arcsec<R<55\arcsec$) gas is
driven strongly radially inwards.  Elsewhere in the disk gas is either stationary or experiences a radially outward torque.  \\

2. We use a unique method to decompose the line-of-sight velocity field of the
CO emission into its radial and azimuthal components. Across the PAWS
field, the sign and magnitude of our derived radial gas streaming motions agree very well with our expectations for how the
molecular ISM should respond to torquing by the non-axisymmetric stellar structure.\\

3. Comparison of the radial torque profile and gas streaming
motions with the molecular gas surface density and star formation rate (as
traced by the combination of H$\alpha$ and 24 $\mu m$ emission), reveals a clear anti-correlation between strong gas flows and active high-mass star formation.  
In M51, radially inflowing gas appears to be less efficiently forming stars.
More generally, gas sitting near the corotation of the disk and the bar or spiral structure forms stars more efficiently than gas that is far from corotation and in motion relative to the background potential.   \\ 

4. We propose that the complementary patterns of gas flow and star
formation in the PAWS field are evidence for the importance of
external pressure on the properties and stability of star-forming
clouds in M51.  More precisely, we develop a simple model based on
the Bernoulli principle that describes changes in cloud surface pressure in the presence of strong streaming motions (dynamical pressure).   
Such changes in surface pressure can make the difference between stability and collapse when, as in M51s spiral arms, clouds are dynamically coupled to their environment (i.e. when the cloud internal and external pressures are similar).  
In M51, cloud stabilization occurs preferentially as the result of strong radial inflow motions in regions of negative torque, but we expect that outflow motions may serve the same role in other systems.   \\

5. Our model successfully reproduces several key observations.  We can retrieve the environment-dependent variation of the slope of the cloud mass
spectrum measured directly from the cloud distribution.  
We can also explain the overall difference in the masses of clouds between the arm and interarm.  Sensitivity to changes in dynamical pressure are also consistent with the observation that not all of M51s clouds are virialized, as evidenced by the lack of a clear size-linewidth relation for clouds in M51.\\

6. We investigate other potential sources of cloud stability (shear, galactic shocks, stellar feedback-driven turbulence) and find that they can not uniformly explain the observed non-monotonic radial dependence of the gas depletion time.   
Cloud formation mechanism also has little influence on the ability of clouds in M51 to collapse and form stars: while cloud collision and agglomeration in the spiral shock are likely present, we find no obvious relation to star formation in M51s current gas reservoir.     \\

7. Although non-circular motions in M51 are particularly high, the influence of dynamical pressure should be common-place in spiral galaxies.  We suggest that gas flows within galaxies are a source of scatter in the Kennicutt-Schmidt star formation relation both within and among galaxies.  
Late-type spiral disks with tightly wound spirals that drive weak streaming motions will form their stars at a faster rate per unit gas mass than early-type disks where streaming motions are higher.   \\

8.  We propose that our model for the dependence of gas depletion time on dynamical pressure links low- and high-z star formation from dense, molecular material.  In particular, we suggest that systems form stars with progressively shorter depletion times the higher their gas fraction.  \\

We thank the IRAM staff for their support during the observations with
the Plateau de Bure interferometer and the 30m telescope.
DC and AH acknowledge funding from the Deutsche Forschungsgemeinschaft (DFG) via grant SCHI 536/5-1 and SCHI 536/7-1 as part of the priority program SPP 1573 'ISM-SPP: Physics of the Interstellar Medium'.
CLD acknowledges funding from the European Research Council for the FP7 ERC starting grant project LOCALSTAR.
TAT acknowledges support from NASA grant \#NNX10AD01G.
During this work, J.~Pety was partially funded by the grant ANR-09-BLAN-0231-01 from the French {\it Agence Nationale de la Recherche} as part of the SCHISM project (\url{http://schism.ens.fr/}).
ES, AH and DC thank NRAO for their support and hospitality during their visits in Charlottesville.
ES thanks the Aspen Center for Physics and the NSF Grant \#1066293 for hospitality during the development and writing of this paper.

\appendix
\section{
A. Mass-based rotation curve for M51}\label{sec:appxRot}  

We use the standard approach to assemble an estimate of the rotation curve in M51 by summing the individual contributions of the stellar and gaseous components to the circular velocity, adopting a simple mass-follows-light model appropriate for spiral galaxies (i.e. the rotation curve can be accounted for with baryons, alone).\footnote{The maximum disk hypothesis has been found to yield good fits inside the optical radius of 3/4 of the galaxies in spiral sample studied by \cite{palunas}.}
\begin{figure}[t]
\begin{center}
\includegraphics[width=.65\linewidth]{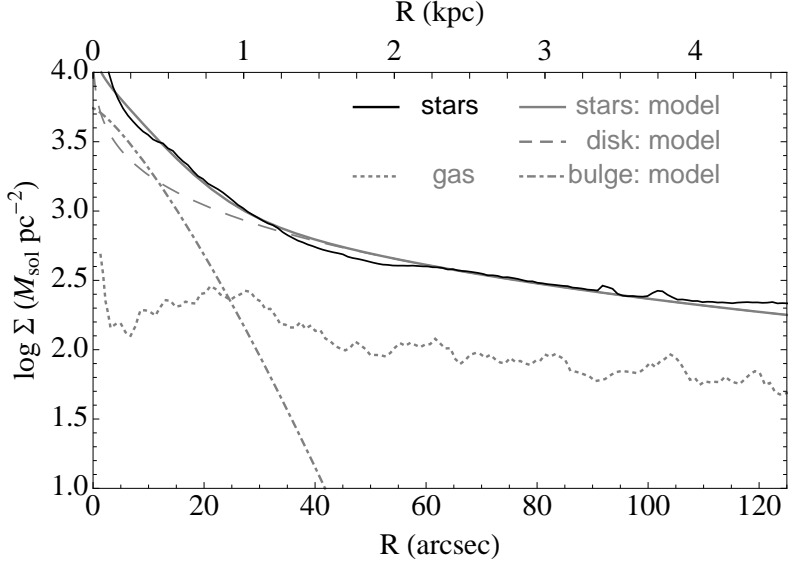}
\caption{Surface density profiles in the central 9 kpc of M51.  The thin gray dotted line shows the total gas surface density $\Sigma_{gas}$=$\Sigma_{H_2}$+$\Sigma_{HI}$.  The solid gray and black lines show the modeled and observed stellar surface density $\Sigma_\star$ assuming the best-fit stellar $M_\star$/$L_{3.6}$ (determined below).  The model components are shown in thin gray lines: exponential stellar disk (dashed) and Gaussian bulge (dot-dashed). }
\label{fig:sigmas}
\end{center}
\end{figure}

This requires first constructing radial profiles of the stellar and gas surface densities.  The total gas surface density $\Sigma_{gas}$ is assembled from measurements of $\Sigma_{HI}$ and $\Sigma_{H_{2}}$ traced by the THINGS HI zeroth-moment map \citep{walter} and PAWS CO(1-0) observations \citep{petyPAWS}, respectively, and includes a factor of 1.36 to account for the presence of Helium.  
As in $\S$ \ref{sec:data} we assume a constant CO-to-$H_2$ conversion factor of $X_{CO}$=2$\times$10$^{20}$ cm$^{-2}$ (K km s$^{-1}$)$^{-1}$ and also include a sky-to-disk plane correction of $\cos{21\degree}$.   

The stellar surface density $\Sigma_{\star}$ is traced by the S$^4$G 3.6 $\mu m$ map of the old stellar light (corrected for non-stellar emission; Meidt et al. 2012).  The (assumed) global 3.6 $\mu m$ mass-to-light ratio $\Upsilon_{3.6}$ (with which $\Sigma_{\star}$=$\Upsilon_{3.6}$$L_{3.6}$$\cos{21\degree}$) is the single free parameter to be fit in comparing the modeled and observed circular velocities.   
The final surface density profiles are shown in Figure \ref{fig:sigmas}.  Below we adopt models of the individual exponential disk and Gaussian bulge components of the stellar surface brightness.    

To estimate the circular velocities associated with the gaseous and stellar disks we adopt the thin disk approximation and model the bulge separately.  The circular velocity at radius $R$ (in the plane of the disk) is given by 
\begin{equation}
V_{c,D}^2(R)=2\pi G R\int_0^\infty \int_0^\infty \Sigma(R') J_0(kR') J_1(kR) R' k dk dR'
\end{equation}
where $J_0$ and $J_1$ are cylindrical Bessel functions (e.g. \citealt{barnesSellwood}; \citealt{barnesSellwood2}). 
Without accounting for disk thickness, which would reduce the circular velocity, the derived $\Upsilon$ is slightly underestimated.  

Following \citet{palunas} we estimate the contribution from the stellar bulge to $V_{c}$ by modeling the bulge component of the stellar surface brightness profile with a Multi-Gaussian expansion
\begin{equation}
\mu_B(a_B)=\sqrt{\frac{f}{\pi}} \Sigma_{k=1}^n \frac{c_k}{R_{B_k}} e^{-a_B^2/R_{B_k}^2}
\end{equation}
where $a_B$=$R^2+(z/(1-\epsilon_B))^2$ is the bulge semi-major axis length,  $\epsilon_B$=1-$q_{p,B}$ is the bulge ellipticity, $q_{p,B}$ is the apparent bulge axis ratio, $f$=$\cos{i}^2+\sin{i}^2/q_B^2$ and, for each of $k$ components, $c_k$ is the total light and $R_{B_k}$ is the scale length.  

This 2D elliptical Gaussian distribution deprojects to a 3D spheroidal distribution with intrinsic axis ratio $q_B$=$(q_{p,B}^2-\cos{i}^2)/\sin{i}^2$
whereby (in the plane of the disk)
\begin{equation}
V_{c,B}^2(R)=4\pi G \Upsilon_{3.6}\sqrt{1-e_B^2}\int_0^R \frac{\mu(R'^2)R'^2 dR'}{\sqrt{R^2-e_B^2R'^2}}
\end{equation}
with $e_B$ the bulge eccentricity, which we take to be zero for simplicity (given the small degree of asymmetry in the central projected light distribution).  We expect this assumption to have less consequence than our requirement for a global $\Upsilon_{3.6}$, uniform from bulge to disk.  
 
\begin{figure}[t]
\begin{center}
\includegraphics[width=.65\linewidth]{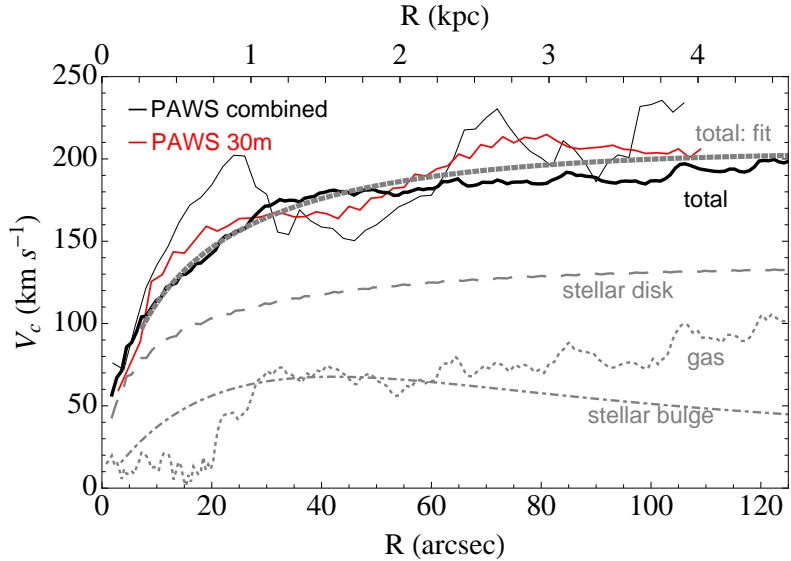}
\caption{Rotation curves in the center ($R$$<$4.5 kpc) of M51.  Circular velocities derived from the PAWS CO(1-0) data are shown in thin lines (combined data, black; single-dish, red; errors are omitted for clarity).  The thick solid black line shows the total $V_c$, calculated as the sum of the individual modeled circular velocities shown in thin gray lines:
stellar bulge component (dot-dashed), stellar disk component (dashed), gas (dotted).  Comparing the red curve with our smooth fit to the total circular velocity (thick gray dotted line) gives a best-fit M/L=0.45$\pm$0.15.  
}
\label{fig:vcs}
\end{center}
\end{figure}

Modeled circular velocities are plotted in Figure \ref{fig:vcs} together with two independent estimates of the rotational velocity measured from the PAWS CO(1-0) observations \citep{colombo2013}.  
The thick black curve shows the rotational velocity measured by fitting tilted-rings to the line-of-sight velocities in the combined interferometric and single dish data (with native 1\arcsec ~resolution; \citealt{petyPAWS}) using the GIPSY task ROTCUR with  fixed center, position angle PA=173$\degree$ and inclination angle $i$=21$\degree$.  
The thick red curve shows the rotational velocity measured from the 30m single-dish data (12\arcsec ~resolution) assuming the same kinematic parameters.    
We expect this curve to be a fairer representation of the true circular velocity, as beam-smearing at lower resolution minimizes the contribution from non-circular streaming motions to the measured rotational velocity (evident as wiggles in the thin curve from the higher resolution data).   Our optimal $\Upsilon_{3.6}$ is chosen based on agreement with this estimate of the circular velocity, as described below.  We adopt a smooth three-parameter fit to the total velocity (i.e. \citealt{fabergall}) as our final model for $V_c$.    
With our mass-follows-light model we find a best-fit stellar $\Upsilon_{3.6}$=0.45 $\pm$0.15. This value agrees well with the value $\Upsilon_{3.6}$=0.5 measured by \citet{eskew}, which is itself very near the value that has been previously assumed to convert between 3.6 $\mu m$ luminosity and mass \citep{leroy08}.  
The maximum disk hypothesis is evidently reliable in this case, at least throughout the PAWS field of view.  

\section{B. Defining the dynamical environment: Present-day Torques }
\label{sec:appxTorques}
The impact of environment is underlined by present-day torques exerted by non-axisymmetric structure in the disk.   We use information supplied by these gravitational torques to define the dynamical environment.  

\subsection{Gravitational torque estimation}
\begin{figure}[t]
\begin{center}
\begin{tabular}{c}
\includegraphics[width=.5\linewidth]{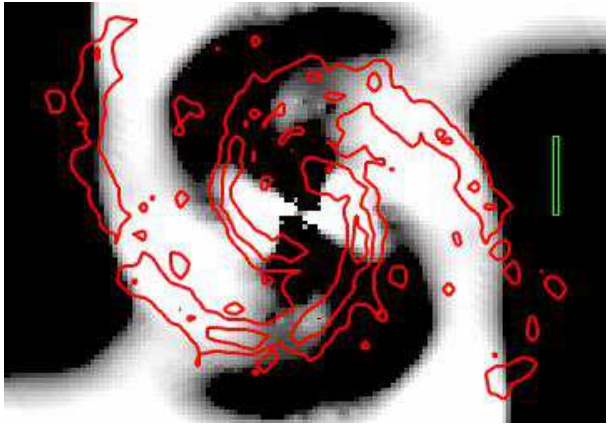}
\end{tabular}
\caption{Map of torques in the inertial frame $R\times\nabla \Phi$, generated from the 3.6 $\mu m$ map of the old stellar light tracing the stellar potential $\Phi$ (see text).  White (black) corresponds to positive (negative) torques that drive motions radially outward (inward).  Contours of the CO intensity are overlayed in red.  The green bar at right indicates 40\arcsec.  ~Image dimensions: $\sim$300\arcsec$\times$200\arcsec. }
\label{fig:maps}
\end{center}
\end{figure}

The S$^4$G 3.6 $\mu m$ map of the old stellar light (Meidt et al. 2012; from which a contribution from non-stellar emission has been removed) presents us with an optimal view of the backbone of present-day gravitational torques.\footnote{Note that the influence of the companion is included to the extent that it has contributed to the reorganization of the stellar mass and structure in the disk mapped in the stellar potential.}  
To generate a map of the stellar potential we first deproject the image according to the inclination $i$=21$\degree$ and major axis position angle PA=173$\degree$ determined by \citet{colombo2013v} from gas kinematics.  We then apply a constant stellar mass-to-light ratio (M/L) to the 3.6 $\mu m$ image and map this into density by assuming a simple tanh vertical distribution with uniform scale height.  
Like the stellar M/L, the adopted vertical distribution will affect the magnitude of the measured torques but not the radial dependence highlighted below, which is our primary concern.  
Triaxiality in the DM halo could potentially introduce a change in the radial behavior, but we expect this to be minimal over our relatively small and centralized field of view covering the inner 9 kpc.  
The dominance of the stellar mass over this area is confirmed by our study of gas kinematics in $\S$ \ref{sec:streaming}, which presents supporting evidence for the conclusions drawn from the stellar potential alone.   

From the potential we calculate a map of the quantity -$R\times\nabla \Phi$ (Figure \ref{fig:maps}) representing the inertial torque per unit gas mass, following \citet{gb05}, \citet{gb09} and \citet{haan}. In the rotating (non-inertial) frame forcing includes a contribution from the `fictitious'  centrifugal and coriolis terms.  
Since the cross product of radius $R$ with these terms go identically to zero in the former case and to zero with averaging over 0 to 2$\pi$ in azimuth in the latter, then a measure of the time average of the rate of change of the angular momentum, the torque, can be obtained by taking azimuthal averages of $R\times\nabla \Phi$, with weighting by the molecular gas surface density.  
This is equivalent to the approach of \citet{gb05} and \citet{haan} where the molecular gas surface density represents the probability of finding the gas at its current position at present.   
Here we use the PAWS observations of CO(1-0) emission to trace the molecular gas surface density assuming a CO-to-H$_2$ conversion factor $X_{CO}$=2$\times$10$^{20}$ cm$^{-2}$ (K km s$^{-1}$)$^{-1}$ \citep{colombo2013}.  
These observations include single dish data, which has been found to be a critical addition to the gas tracers used for this purpose (\citealt{gb05}; \citealt{vdlaan}); extended emission on the largest scales must be included for accurate determination of the net torque.  
Note that the assumed conversion factor is another potential influence on the magnitude of the derived torques.  For this reason in the upcoming section we present the torque profile in units of the net torque over the PAWS field of view. 

\subsection{Inventory of distinct dynamical environments\label{sec:env}}
Figure \ref{fig:torqsfe} in the main text shows the radial profile of azimuthally averaged torques across the PAWS field of view.  Each crossing from negative to positive marks a corotation radius (CR), while crossings from positive to negative torque coincide with a switch in predominance to a new pattern.  Three dynamically distinct zones are identified, as described below.  

The first CR in Figure \ref{fig:torqsfe} occurs just outside the ``butterfly" pattern in the central 20\arcsec ~of the torque map, which is characteristic of the bar influence (i.e \citealt{gb}).  Comparison with the velocity field confirms that the gas is responding to torquing by the stellar bar (first identified in the NIR by \citealt{zrr}), even though the stellar bar lacks a molecular counterpart.   
The bar is responsible for the +10-15$\degree$ twist in the orientation of the line-of-nodes, previously interpreted as evidence for variation in the galaxy major axis tracing a warp or twist of the disk \citep{shetty}.   This marks the first evidence that the bar, rather than the spiral, dominates the kinematics in the central 30\arcsec.   

The influence of the bar also extends to gas morphology, as evidenced by the build-up of a molecular ring at the overlap with the molecular spiral arms just exterior, near R=35\arcsec.  
Opposing torques from inside the ring, where the bar drives gas outwards, and outside the ring, where torques drive gas inwards (see below), result in the pile-up of gas at the location of the ring.   

Beyond the second corotation at $R_{CR}$= 55\arcsec ~gas is again driven outwards, across the outer zone of the CO-bright spiral arms.  
Radial outflow continues until the start of the third distinct pattern, near $R$=85\arcsec, where the direction of flow is again reversed.  This location is consistent with previous estimates for the start of the distinct outer, material spiral pattern (e.g. \citealt{tully}; \citealt{vog93}; \citealt{meidt08b}) which continues beyond the edge of our field of view.  

In total, the gravitational torques expose three dynamically distinct zones dominated by at least three unique non-axisymmetric structures.  The number, multiplicity and radial domains of patterns present in the disk will be the subject of a more detailed study in the future \citep{colombo2013v}.  Below we briefly review supporting evidence for our identification of these three main environments.  
\begin{figure}[t]
\begin{center}
\includegraphics[width=.65\linewidth]{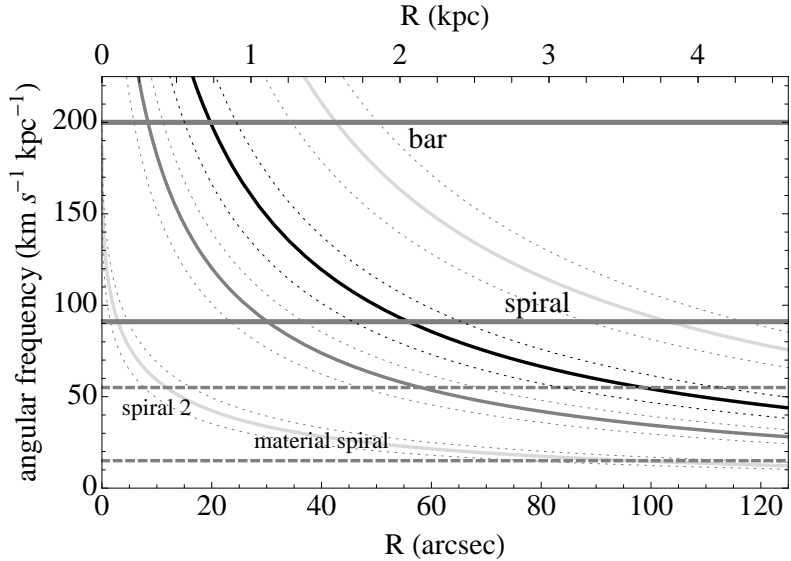}
\caption{Pattern speed estimates for structure in the disk of M51.   Angular frequency curves are shown: $\Omega$ (Black), $\Omega-\kappa/4$ (Gray), $\Omega\pm\kappa/2$ (Light Gray).  Estimates for the bar and spiral pattern speeds implied by the corotation radii revealed by the torque profile in Figure \ref{fig:torqsfe} are shown as gray horizontal lines.  A possible second spiral speed (see text for discussion) and the rotation of the material arms are also indicated in gray horizontal dashed lines.
\vspace*{.25in}}
\label{fig:speeds}
\end{center}
\end{figure}

\subsubsection{Complimentary evidence for multiple patterns\label{sec:complimentary}}
We expect at least three distinct pattern speeds in the central 9 kpc of M51, coinciding with three distinct structures: the bar, the main spiral and an outer spiral.    
For the main spiral pattern with corotation at $R_{CR}$=55\arcsec ~we estimate a pattern speed of $\Omega_p$$\sim$90 km s$^{-1}$ kpc$^{-1}$ based on intersection with our adopted angular frequency curve (see Fig. \ref{fig:speeds}).  
This is very close to the value measured by \cite{meidt08b} with the radial Tremaine-Weinberg (TWR; \citealt{mrm}, \citealt{meidt08a}) method using lower-resolution molecular gas observations as a kinematic tracer, and assuming a similar major axis position angle of PA=175$\degree$.\footnote{Accounting for differences in adopted distance and inclination angle}$^,$\footnote{Two speeds from an alternatively plausible TWR solution, which assumes a lower PA=170$\degree$, straddle our adopted value $\Omega_p$$\sim$90 km s$^{-1}$ kpc$^{-1}$.  
Based on evidence here for the bar influence on morphology and molecular gas kinematics (see $\S$ \ref{sec:streaming}), though, it seems likely that the higher inner pattern speed measured by \cite{meidt08b} reflects a combination of the speed at $\Omega_p$=90 km s$^{-1}$ kpc$^{-1}$ and a yet higher bar pattern speed.  This circumstance has been encountered in other galaxies (\citealt{meidt09}), where the resolution in the TWR solution was limited by that of the kinematic tracer.  
According to our estimate of the bar CR, the bar zone would have been probed by only the inner three to four $\sim$6\arcsec ~radial bins used by \cite{meidt08b}.  
In addition, given the orientation of twisted line-of-nodes manifest by the bar, the higher PA=175$\degree$ major axis orientation for the TWR solution presumably disfavors the detection of the bar influence, more so than for a lower assumed PA;  as for the TWR solution at PA=170$\degree$ a third solution with PA=165$\degree$ measured by \cite{meidt08b}  suggests an inner pattern speed $>$90 km s$^{-1}$ kpc$^{-1}$. 
}

At smaller radii, the gravitational torques suggests a transition to a distinct bar pattern, with $R_{CR}$=20\arcsec, which is comparable to the location recently suggested by \cite{zhangbuta2012}.  From this we estimate a bar pattern speed of $\Omega_b$$\sim$200 km s$^{-1}$ kpc$^{-1}$.   In this case, the bar and spiral pattern speeds offer a physically plausible scenario, as the corotation of the bar coincides with the inner 4:1 resonance of the spiral.  
Resonance overlaps of this kind have been identified in barred spiral simulations (\citealt{rs}; \citealt{deb06}; \citealt{minchev}) and may be characteristic of nonlinear mode coupling (e.g., \citealt{tagger} and \citealt{syg}), whereby energy and angular momentum are transferred between the modes.  Note that with this set of pattern speeds, the molecular ring sits near the spiral's inner 4:1 resonance (and near the bar's outer 4:1 resonance) and forms at the intersection of positive torquing by the bar outside its CR with the negative torquing by the spiral inside its CR.  The inner 4:1 resonance has previously been suggested to favor the population of gas in the form of such a ring \citep{gb09}.  
 (We also note that the coincidence of the end of the bar with its CR makes this bar a so-called `fast' bar.)    

This same process could be at work further out in the disk, in the zone between 35\arcsec ~and 70\arcsec.   At its $R_{CR}$=55\arcsec  ~the speed $\Omega_p$=90 $\kmskpc$ could overlap with the inner 4:1 resonance of a lower speed of $\Omega_{p,2}$$\sim$55 $\kmskpc$ (see Figure \ref{fig:speeds}), just before the start of the material arms.  
This second spiral speed falls just within the range of speeds suggested for a similar second speed in the TWR solution at 170$\degree$ measured by Meidt et al. (2008).   

The existence of this second more slowly rotating pattern may be plausible for several reasons: 1) the gas kinematics less clearly show the pattern of outflow in the post-corotation zone 55\arcsec$<$$R$$<$70\arcsec ~as would be expected for a pattern outside its CR.\footnote{The radial torque profile $<$$\Gamma$$>$($R$) (Figure \ref{fig:torqsfe}) 
suggests that radial gas inflow stops at $R_{CR}$=55\arcsec, rather than continue until $R$=70\arcsec, as suggested by the radial streaming motions $v_r$, measured in Appendix \ref{sec:appendixStream}.  
This difference could be alleviated with a constant offset to either $v_r$ or $<$$\Gamma$$>$ (with an impact on the locations of the other CR radii, as well), or it could indicate a more complicated gaseous response to the torquing.}     
Two patterns, each propagating inside their CR would maintain radial inflow through this zone.   2) The spatial offset between young star and gas tracers does not clearly switch sign across the corotation at $R_{CR}$=55\arcsec ~as predicted in response to a spiral density wave around CR.   
Again, two patterns, both propagating inside CR, and both driving gas radially inwards, removes the possibility that young stars will be found upstream of the gaseous spiral, as otherwise expected for a single pattern outside its CR.   3) 
While the 3.6 $\mu m$ map of old stellar light shows a pronounced $m$=2 perturbation, this is identifiable as a two-armed spiral only outside $R$$\sim$55\arcsec.  Inside this radius, in the zone 35\arcsec$<$$R$$<$55\arcsec, ellipse fits to 3.6 $\mu m$ isophotes reveal a more oval, bar-like structure (see also \citealt{schinner2013}).  
Other stellar traces (from FUV to H$_\alpha$) reveal a similarly marked absence of a clear spiral pattern in this zone.  In this case, the higher pattern speed of $\Omega_p$$\sim$90 $\kmskpc$ might apply to a bar terminating at its corotation radius, while the lower pattern speed $\Omega_{p,2}$$\sim$55 $\kmskpc$ could describe the strong spiral pattern.  
We might then interpret the gaseous response to the oval perturbation, which is very definitely in the form of a two-armed spiral, as equivalent to dust lanes running along the length of a stellar bar.  
Alternatively, the main spiral pattern could overlap with a lower, outer pattern with speed $\Omega_p$=20 $\kmskpc$ near where its OLR and the outer pattern's ILR coincide (see Figure \ref{fig:speeds}).  

For simplicity in what follows, we will describe the dynamical environment by three distinct zones dominated by an inner, or nuclear, bar and two (inner and outer) molecular spirals.  As explored in $\S$ \ref{sec:streaming}, the observed gas motions reflect the predicted combination of radial inflow and outflow that, in the spiral arm region, correspond to flow along and through the arm.  

\section{C. Decomposing the PAWS line-of-sight velocity field}
\label{sec:appendixStream}  
\begin{figure}[t]
\includegraphics[width=.51\linewidth]{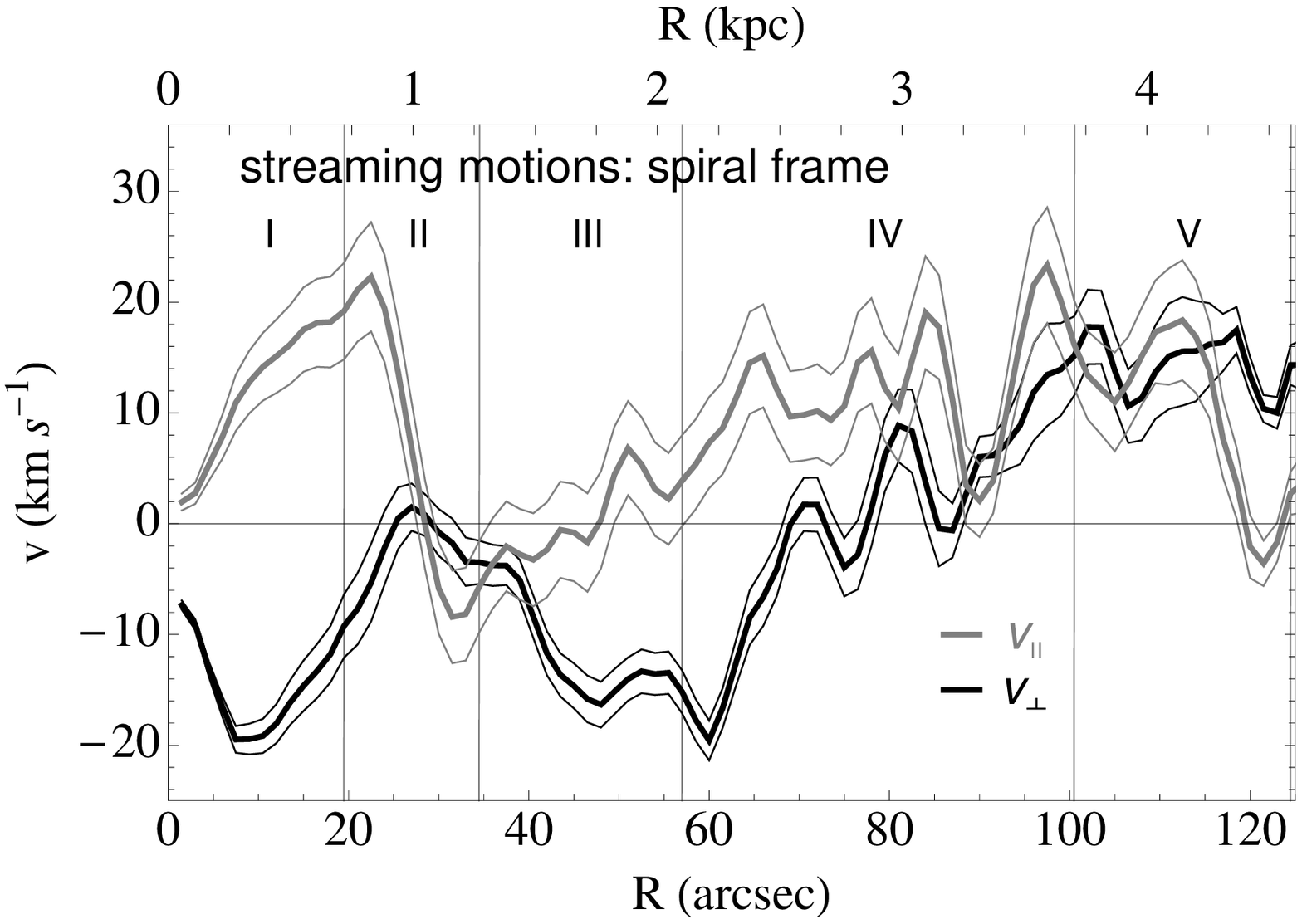}
\includegraphics[width=.51\linewidth]{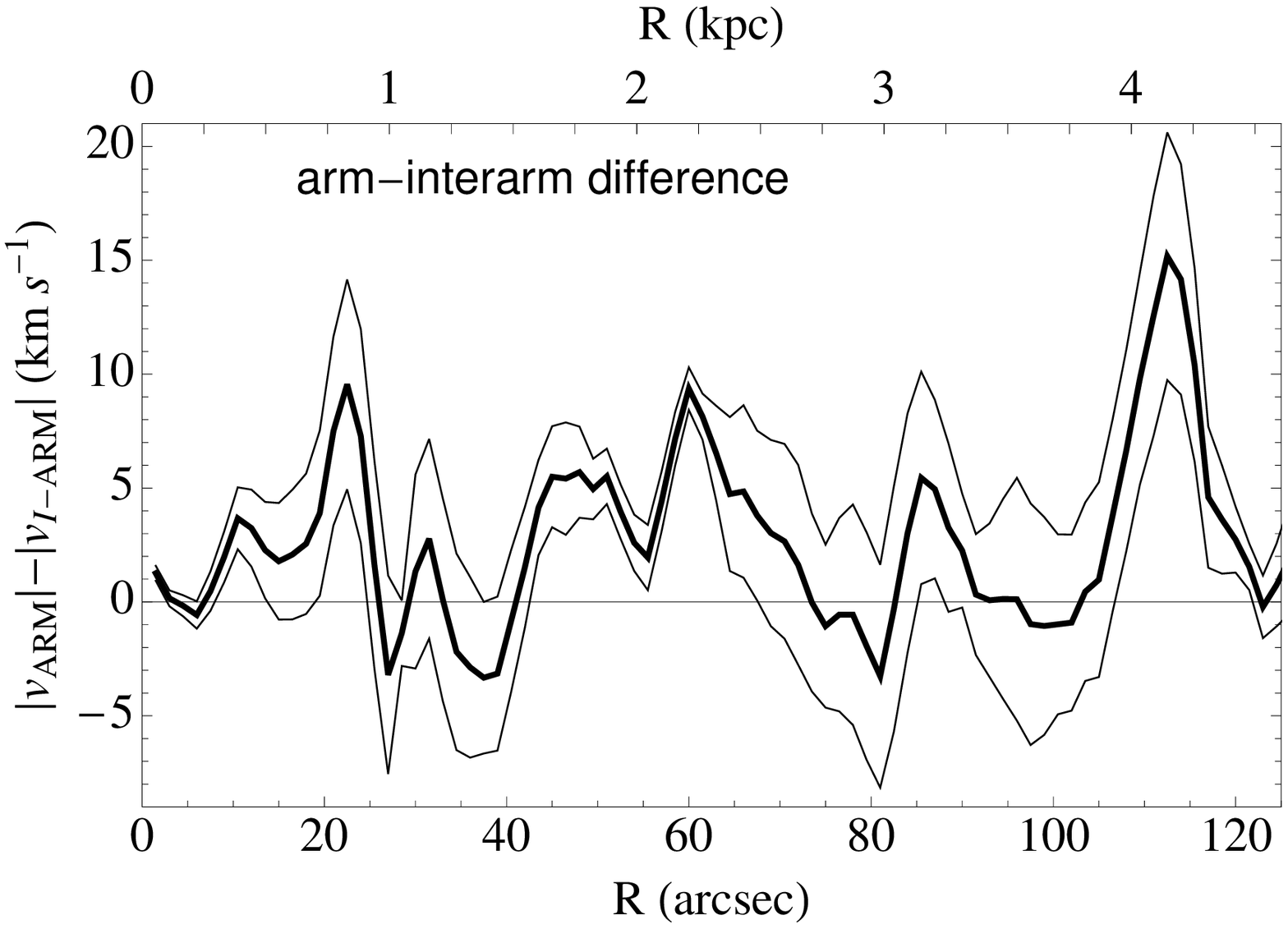}
\caption{(Left) CO intensity-weighted velocities parallel ($v_\parallel$, blue) and perpendicular ($v_\perp$, red) to the spiral arm, calculated in radial bins as described in the text from the reconstructed radial and azimuthal velocities (see Figure \ref{fig:stream} in the main text).  
Dashed lines represent the uncertainty associated with each velocity component, calculated from the errors shown in Figure \ref{fig:stream} (i.e. the rms dispersion in solutions with $\pm$5$\degree$ variation in the assumed spiral arm pitch angle $i_p$ and the uncertainty introduced by our adopted rotation curve).  
The intensity-weighting focuses our measurements on the CO-bright spiral arms.  (Right) The difference in the magnitude of non-circular streaming motions in the arm and interarm.  }
\label{fig:spiralframe}
\end{figure}

The nearly logarithmic
spiral in M51, spanning over 5 kpc in radius, presents a unique
opportunity to decompose the line-of-sight velocity
\begin{eqnarray}
V_{los}(R,\theta)&=&V_{sys}\nonumber\\
&+&[(V_c(R)+v_\phi(R,\theta))\sin{(\theta)}\nonumber\\
&+&v_r(R,\theta) \cos{(\theta)} ]\sin{i}\nonumber\\
\end{eqnarray}
in to its radial and azimuthal components $v_r$ and $v_{\phi}$.  Here $V_{sys}$ is galaxy systemic velocity, $V_c$ is the disk circular velocity, and $\theta$ is measured with respect to the line of nodes.  
Following \citet{colombo2013v}, we adopt an inclination of $i$=21$\degree$ and a major axis position angle of $\theta_{PA}$=173$\degree$.

The key to decomposing $V_{los}$ is the idealized arm frame in which streaming motions $v_r$ and $v_{\phi}$ will be nearly constant in small segments at constant arm phase $\psi$ parallel to each arm.  
Adopting a spiral arm pitch angle of $i_p$=21$\degree$ we transform from cartesian to log-polar coordinates and define a grid of rectangular segments or bins aligned with the spiral arm.  
In each bin we solve for $v_r$ and $v_{\phi}$ via simple matrix inversion, according to 
 \begin{eqnarray}
(V_{los}-V_{sys})/\sin{i}&&\nonumber\\
-V_c\sin(\theta)&=&v_r(\psi) \cos(\theta)\nonumber\\
&+&v_\phi(\psi) \sin(\theta),\label{eq:vsoln}
\end{eqnarray}
where the coefficients are specified by the geometric factors.  

We calculate maps assuming three different pitch angles: a nominal $i_p$=21$\degree$ (i.e. \citealt{shetty}) and a second set offset by $\pm$5$\degree$.  The average of the solutions from these three cases define our estimate for each component, while the dispersion in these values supplies a measure of our uncertainty, together with the uncertainty introduced by our adopted rotation curve (see Appendix \ref{sec:appxRot}). 
To accommodate for real variation in the molecular spiral arm pitch angle (determined to be 5$\degree$ between the arms and as much as 20$\degree$ along the arms by \citealt{patrikeev}), we size the spiral segments over which the solution is calculated to 5.6$\degree$ wide and $\sim$$\log{r (\arcsec)}$=0.011.  
These segments cover a smaller area toward the center, so the solution in the bar zone, where a logarithmic-polar deprojection may be less fitting, should nevertheless be accurate. 

Prior to solution the contribution of the circular velocity to the observed l.o.s. velocity field is removed (see eq. \ref{eq:vsoln}).   We adopt the circular velocity estimated from our 1D mass model of M51, constructed as described in Appendix 1.  
By using a mass-based reconstruction of the circular velocity we avoid the uncertainties typically inherited by circular velocity estimates made directly from the l.o.s. velocity field (e.g., with tilted-ring fits), namely the contribution of azimuthal streaming motions.  
The result is a much more smooth, physically motivated model of the circular velocity curve that should insure accurate streaming motion solutions.  
We use the uncertainty introduced in the rotation curve by our choice of stellar $M_{\star}/L_{3.6}$=0.45$\pm$0.15 (Appendix 1) to define the error in our measured streaming velocities, together with the measured pitch angle uncertainty.  
Both are typically $<$25\%.  

The tangential and radial streaming motions (shown in Figure \ref{fig:stream} in the main text) translate into motion directed along and through the arm, channeled by the stellar spiral.  
This is depicted more clearly in the top left of Figure \ref{fig:spiralframe}, showing intensity-weighted averages of the velocities in the spiral arm frame, assuming $i_p$=21$\degree$.  Here 
\begin{eqnarray}
v_\perp&=&v_r \cos(i_p)+v_{\phi}\sin(i_p)\nonumber\\
v_\parallel&=&-v_r \sin(i_p)+v_{\phi}\cos(i_p)
\end{eqnarray}
where $v_\perp$ runs perpendicular to the arm and $v_\parallel$ runs parallel to the arm. Note that in the zone 40\arcsec$<$$r$$<$60\arcsec, the velocities $v_\perp$ perpendicular to the spiral arm arise almost completely with the radial motions since the azimuthal streaming in this zone is at a minimum.  
The spiral arm pitch angle is meanwhile increasing here, by up to 30$\degree$, so that the radial streaming motions could be distributed slightly more in the direction parallel to the spiral.  
This would lead to larger $v_\parallel$ than measured here (assuming $i_p$=21$\degree$), although the perpendicular motions still dominate in this case. 

As described in $\S$ \ref{sec:streaming}, we use CO intensity as a weighting factor in binned azimuthal averages to reveal the velocities characteristic of the (CO bright) spiral arms.  Profiles with inverse-intensity weighting, which highlight the inter-arm streaming velocities, are qualitatively similar but show velocities $\approx$ 5-10 km s$^{-1}$ lower than the profiles in Figure \ref{fig:stream}.  The difference in the magnitudes of streaming motions in the arm and interarm is shown in the right panel of Figure \ref{fig:spiralframe}.

\section{D. Gravitational Disk Stability}
\label{sec:appxCritdens}
In this section we examine in greater detail the stability of the 
molecule-dominated gas disk in M51.  We quantify and compare the critical surface density for dynamical pressure to that in the presence of shear, Coriolis forces and
tidal forces, three sources of cloud stability of potential consequence for the global pattern of star formation.   

Following \cite{toomre} we take 
\begin{equation}
\Sigma_{toomre}=\frac{\alpha \sigma\kappa}{\pi G}
\end{equation}
as defining the threshold for stability in a thin rotating gas disk, whereby the Toomre $Q$=1 when $\Sigma_{toomre}=\Sigma_{gas}$.   Here $\sigma$ is the gas velocity dispersion, $\kappa$ is the epicyclic frequency measured from our rotation curve model (see Appendix \ref{sec:appxRot}) and $\alpha$ is a dimensionless factor calibrated empirically.\footnote{
Values $\alpha\neq$1 are chosen in order that $Q$=1 at the location of the observed star formation threshold.   
As suggested by \cite{schaye} and \cite{dBW} deviations in $\alpha$ from unity may be commonly required only because measurements of the gas velocity dispersion from observations of the atomic phase systematically overestimate the true $\sigma$ of gas in a cold phase, which we can better approximate with  the PAWS observations 
(as confirmed in M51 by \citealt{colombo2013v}).  
}  (In the following we let $\alpha$ = 1 but also discuss the impact of variation in this value.)   Below this threshold, Coriolis forces spin up the gas so that centrifugal forces counter the self-gravity of the gas, resulting in a suppression of cloud (and star) formation.   Above the threshold, gas is unstable to large-scale collapse, leading to star formation.  

The threshold for cloud stability in the presence of shear, as derived by \cite{elmshear} (see \citealt{luna}), can be written
\begin{equation}
\Sigma_{shear}=\frac{2.5\alpha_A\sigma A}{\pi G}
\end{equation}
where $A$ is the Oort parameter and again with a dimensionless factor $\alpha_A$, which we set to unity.  We consider separately the background shear due to differential rotation and the spiral arm shear, accounting for gradients in the azimuthal spiral arm streaming in the definition of $A$ (in addition to background shear), as in $\S$ \ref{sec:stabShear} in the main text.   

Alternatively, cloud shearing via tidal forces will become ineffective above the threshold
\begin{equation}
\Sigma_{tide}=\frac{\sigma [3A(A-B)]^{1/2}}{\pi G}
\end{equation}
following \cite{kenney1993}, with Oort $B=\Omega-A$.   Note that these formulations of cloud stability ignore any modulation/dissipation of shear, tidal or Coriolis forces due to additional effects, such as magnetic fields or viscosity.

According to eq. (5), the critical surface density in the presence of dynamical pressure can be written
\begin{equation}
\Sigma_{DP}=\frac{\sigma^2}{\pi GH}e^{v_S^2/4\sigma^2}\approx\frac{\sigma\Omega}{\pi G}e^{v_S^2/4\sigma^2} \label{eq:critDP}
\end{equation}
letting the characteristic surface density of virialized clouds $\Sigma_{vir}$=$\sigma^2/\pi GH$, and using that the gas scale height $H$ is approximately $\sigma/\Omega$ near to hydrostatic equilibrium in self-gravitating gas.  

\begin{figure}
\begin{centering}
\begin{tabular}{cc}
\includegraphics[width=.51\linewidth]{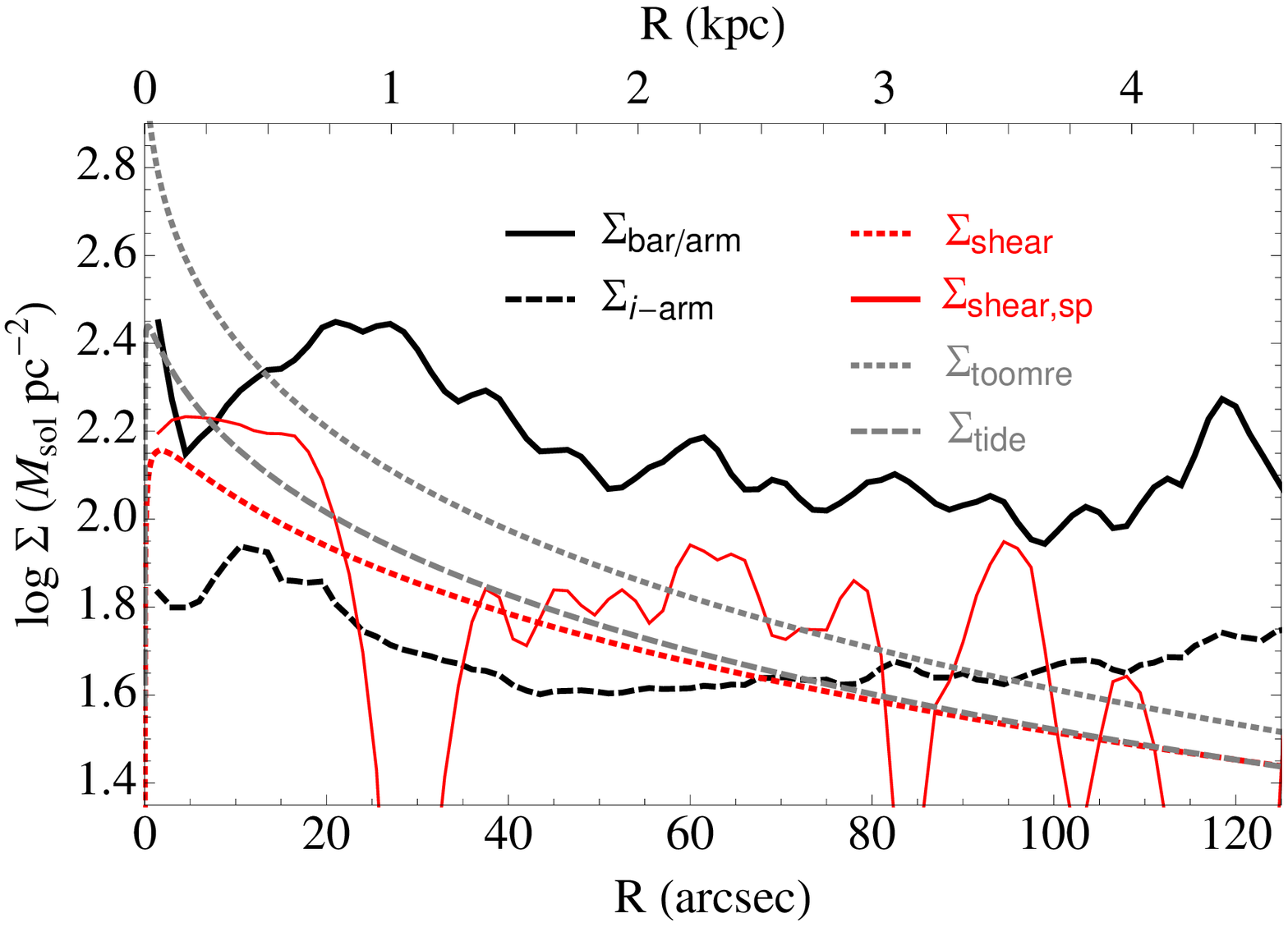}&\includegraphics[width=.51\linewidth]{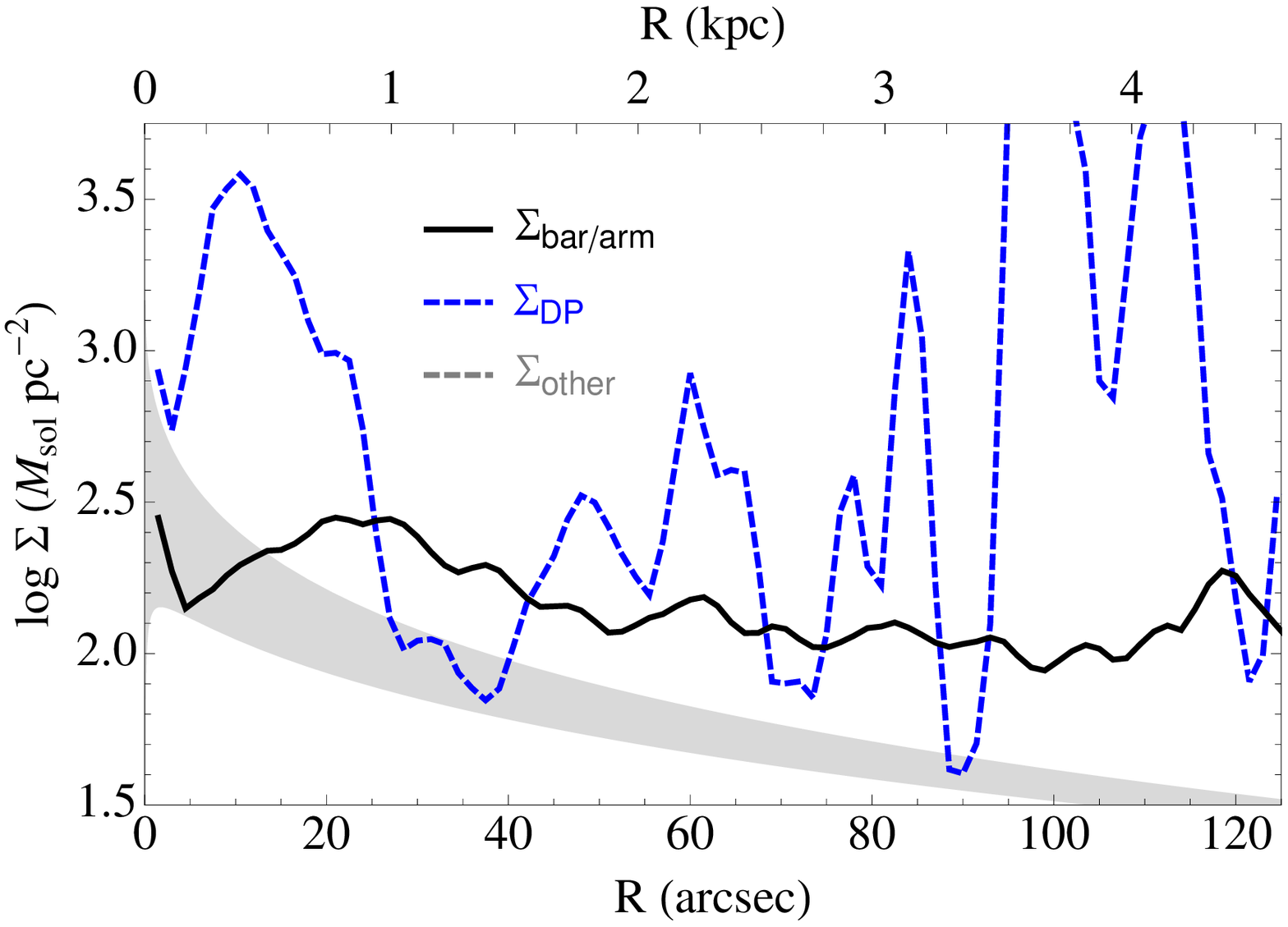}
\end{tabular}
\end{centering}
\caption{\small (Left) Comparison of the observed bar/arm ($\Sigma_{bar/arm}$; solid black) and interarm ($\Sigma_{i-arm}$; dashed black) molecular gas surface densities in M51 with the critical surface densities for stabilization against gas self-gravity via coriolis forces (the so-called Toomre critical surface density $\Sigma_{toomre}$; gray dotted), tidal forces ($\Sigma_{tide}$; dashed gray) and shear ($\kappa$ replaced with Oort parameter $A$) in the disk ($\Sigma_{shear}$; red dotted) and in the arm ($\Sigma_{shear,sp}$; red solid).  
(Right) Comparison between the critical surface densities for shear, tidal and Coriolis forces ($\Sigma_{other}$; shown together as a gray band) with the surface density threshold for stabilization via dynamical pressure ($\Sigma_{DP}$; dashed blue).  The observed bar/arm surface density ($\Sigma_{bar/arm}$; solid black) is repeated from the left panel.  
}
\label{fig:critdens1}
\end{figure}

In Figure \ref{fig:critdens1} we compare the radial profiles of this set of critical surface densities with the observed arm and inter-arm gas surface densities.  As shown on the left, the shear critical surface density $\Sigma_{shear}$ falls below the observed surface density in
the arm almost everywhere, suggesting that shear is not responsible for
destroying clouds in the arm, consistent with the argument of, e.g., Elmegreen
et al. (2003).  The reverse appears to be true in the interarm where, as considered
by \cite{colombo2013}, the reduced cloud number density (compared to the arm) reflects an efficient cloud destruction mechanism. 

In the zone of the bar, the shear and
Toomre critical surface densities approach and even exceed the observed surface density, lending 
support to the idea that Coriolis forces plus shear (and shocks) in the bar
can stabilize clouds and prevent subsequent star formation.  However, several
lines of evidence suggest that cloud stabilization through this more traditional shear/shock scenario in the bar is
weak and likely ineffective: 1) the bar itself is weak,
as indicated by its small size and oval shape, 2) there is little evidence
for a clearly defined pair of dust lanes running along the length of the bar
(traced by either molecular gas or optical extinction), which are
characteristic of a strong bar shock  and 3) as shown on the right of Figure \ref{fig:critdens1}, the threshold for stabilization through dynamical pressure is higher than these other thresholds, exceeding $\Sigma_{gas}$ by as much as an order of magnitude.  
Indeed, the critical surface density for dynamical pressure as expressed in eq. (\ref{eq:critDP}) will always exceed the traditional Toomre critical surface density where $v_S$$>$2$\sigma\sqrt{\ln{\delta}}$ for $\delta=\kappa/\Omega$ (i.e. $\delta=\sqrt{2}$,1 or 2 for rising, flat or Keplerian rotation curves, respectively.)  

\begin{figure}
\begin{centering}
\begin{tabular}{c}
\includegraphics[width=.51\linewidth]{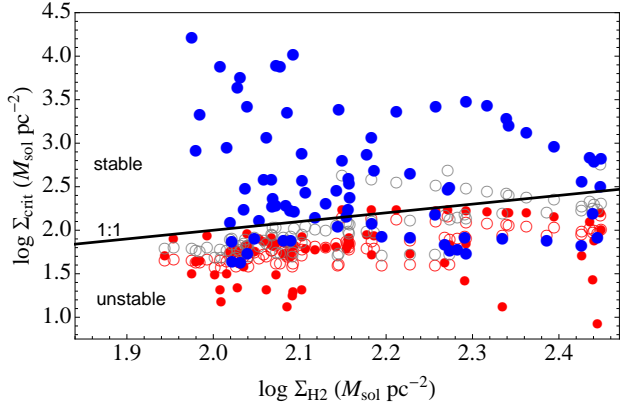}
\end{tabular}
\end{centering}
\caption{\small Critical surface densities vs. $\Sigma_{H2}$ in the bar and spiral arms where different symbols and colors depict the potential for stabilization against gas self-gravity via Coriolis forces (the so-called Toomre critical surface density $\Sigma_{toomre}$; gray open symbols), shear due to differential rotation ($\Sigma_{shear}$; open red) and including non-circular motions ($\Sigma_{shear,sp}$; solid red) and dynamical pressure (solid blue).  Measurements are extracted in 1.5\arcsec ~bins from the radial profiles in Figure \ref{fig:critdens1}.    The solid black line shows the line of stability.  
}
\label{fig:critdens2}
\end{figure}

According to Figure \ref{fig:critdens1}, it appears that star formation in the bar zone may be suppressed and related most directly to
stabilization through dynamical pressure.  
Stabilization via dynamical pressure is also evidently the strongest (if not the only) interpretation for the suppression of star formation along specific portions of spiral arms.  Only $\Sigma_{DP}$ exceeds the observed arm surface densities at radii 35\arcsec$\lesssim$$R$$\lesssim$60\arcsec, where gas depletion times are measured to be longest (Figure \ref{fig:torqsfe}), as well as at larger radii (i.e. $R$$\sim$80\arcsec ~and $R$$\sim100$\arcsec, where $\tau_{dep}$ is also high.)  As plotted in Figure \ref{fig:critdens2}, dynamical pressure is capable of supplying stability where other mechanisms fall short.  

Note that, even if with adjustments to the empirical calibration factors $\alpha$ and $\alpha_A$, cloud stabilization through Coriolis forces and shear cannot account for the differences in star formation properties between neighboring radial zones.  Only dynamical pressure entails a non-monotonic radial variation in gas stability, given the pattern of streaming motions.  



\end{document}